\documentclass[a4paper,onecolumn,11pt]{quantumarticle}
\pdfoutput=1
\usepackage[utf8]{inputenc}
\usepackage[english]{babel}
\usepackage[T1]{fontenc}
\usepackage{amsmath}
\usepackage{hyperref}
\usepackage{tikz}
\usepackage{lipsum}
\usepackage{dcolumn}
\usepackage{bm}
\usepackage{amsmath,amssymb}
\usepackage{xcolor}
\usepackage[utf8]{inputenc}
\usepackage{amsmath,amssymb}
\usepackage{graphicx}
\usepackage{quantikz}
\usepackage{bbm}
\usepackage{dcolumn}
\usepackage{enumerate}
\usepackage{amsthm}
\usepackage{mathrsfs}
\usepackage{float}
\usepackage{caption} % 关键：用于自定义标题格式
\usepackage{subcaption} % 子图支持（但需禁用自动标签）
\usepackage{tikz}
\usetikzlibrary{3d,arrows.meta}
\usepackage{graphicx}
\usepackage{booktabs} % 用于美观表格线
\usepackage{array} % 增强表格功能
\usepackage{setspace}
\usepackage{amsmath,amssymb} % For mathematical symbols and equations
\usepackage{graphicx} % For including figures
\usepackage[numbers]{natbib} % For bibliography management
\usepackage{hyperref} % For hyperlinks in the PDF
\usepackage{lipsum} 
\newtheorem{theorem}{Theorem}
\newtheorem{proposition}{Proposition}
\newtheorem{lemma}{Lemma}
\begin{document}
	
	\title{Quantum State Preparation by Improved MPS Method}
	\author{Chao Wang}
	\affiliation{Origin Quantum Computing,  Hefei,  China}
	\author{Pengrui Zhou}
	\affiliation{Origin Quantum Computing,  Hefei,  China}
	\author{Xi-Ning Zhuang}
	\affiliation{CAS Key Laboratory of Quantum Information, University of Science and Technology of China, Hefei, 230026, China}
	\author{Ziwei Cui}
	\affiliation{Origin Quantum Computing,  Hefei,  China}
	\author{Menghan Dou}
	\affiliation{Origin Quantum Computing,  Hefei,  China}
	\author{Zhao-Yun Chen}
	\email{chenzhaoyun@iai.ustc.edu.cn}
	\affiliation{Institute of Artificial Intelligence, Hefei Comprehensive National Science Center}
	\author{Guo-Ping Guo}
	\email{gpguo@ustc.edu.cn}
	\affiliation{Origin Quantum Computing,  Hefei,  China}
	\affiliation{CAS Key Laboratory of Quantum Information, University of Science and Technology of China, Hefei, 230026, China}
	\affiliation{Institute of Artificial Intelligence, Hefei Comprehensive National Science Center}
	
\begin{abstract}
Efficient encoding of classical information plays a fundamental role in numerous practical quantum algorithms. However, the preparation of an arbitrary amplitude-encoded state has been proven to be time-consuming, and its deployment on current noisy devices can be challenging. In this work, we propose an improved Matrix Product State(MPS) method protocol with an exponential reduction on the circuit depth, as well as topological adaptability. By refined utilization of the disentangling principle, we also reduce approximately 33\% two-qubit gate count. To validate our method, we study various families of functions and distributions with provably bounded MPS rank. Numerical experiments show that our method significantly reduces circuit depth while achieving higher fidelity for states arising in financial and other applications.
\end{abstract}
	\maketitle
	\section{Introduction}
	
	Efficient quantum state preparation is a fundamental requirement for quantum computation, enabling algorithms with proven or potential speedups~\cite{harrow2009quantum,childs2017quantum,martin2021toward,childs2020quantum,an2023linear,berry2024quantum,fang2023time}. This task is particularly challenging in the Noisy Intermediate-Scale Quantum (NISQ) era~\cite{johnson2012heralded, takita2017experimental, kjaergaard2020superconducting, song2019quantum, gyongyosi2019quantum}. NISQ devices are characterized by short coherence times and high error rates, especially for two-qubit gates, which are the primary bottleneck for circuit execution~\cite{barratt2021parallel, ge2024quantum, camps2022algebraic}. Consequently, the viability of a state preparation method is largely determined by the depth of the resulting quantum circuit. While powerful, general-purpose techniques such as Quantum Singular Value Transformation (QSVT)~\cite{gilyen2019quantum, mcardle2022quantum, guo2024nonlinear} or other arbitrary state preparation algorithms~\cite{plesch2011quantum, marin2023quantum, gonzalez2024efficient} exist, their implementation requires deep circuits with a vast number of two-qubit gates, making them impractical for current hardware. This challenge is exacerbated by the overhead from decomposing multi-controlled gates and the constraints of limited qubit connectivity on physical chips~\cite{abughanem2025ibm}. Additionally, training-based approaches~\cite{zhuang2024statistics, benedetti2019generative, zoufal2019quantum, zhu2022generative}, though flexible and expressive, often suffer from difficulties in estimating circuit depth and quantum resource requirements, further limiting their applicability on NISQ platforms. As an alternative, Matrix Product State (MPS) based methods have emerged as a promising direction. By representing the target state as an MPS, one can construct quantum circuits whose depth scales linearly with the number of qubits, representing a more feasible approach for state preparation on near-term devices~\cite{wang2017simulations, huggins2019towards, ran2020encoding, holmes2020efficient, melnikov2023quantum}.
	
The MPS method provides a structured, resource-efficient framework for quantum state preparation, particularly for states exhibiting area-law entanglement common in physical systems~\cite{cirac2021matrix, schollwock2011density, zhang2022qubit}. Conventional MPS-based methods map the tensor network structure directly onto a quantum circuit, typically yielding a circuit depth of $O(n)$ on hardware with chain or ring connectivity~\cite{iaconis2024quantum, ran2020encoding, dborin2022matrix}. While this linear scaling is a significant improvement over general-purpose methods, it still presents a scalability challenge as qubit counts increase and circuits become too deep, with many unused time slots, to execute within coherence times. Recognizing this limitation, recent theoretical works have shown that logarithmic~\cite{malz2024preparation} or even constant-depth~\cite{smith2024constant} preparation is possible for MPS states. However, these advanced schemes often involve significant practical trade-offs: they may rely on non-unitary operations, mid-circuit measurements, and numerous ancilla qubits, which are costly and error-prone on NISQ platforms. Furthermore, their non-unitary nature makes them difficult to use repeatedly in amplitude amplification or estimation circuits in algorithms such as quantum Monte Carlo integration~\cite{rebentrost2018quantum,stamatopoulos2020option,gonzalez2023efficient,agliardi2023quadratic}. Therefore, a critical open challenge is to develop a unitary state preparation scheme that achieves logarithmic circuit depth—the established optimum for unitary methods~\cite{smith2024constant, malz2024preparation}—while remaining compatible with the practical constraints of quantum hardware.
	
In this work, we address these challenges by introducing an IMPS preparation scheme that achieves logarithmic circuit depth (without considering chip architecture) and is optimized for hardware with topological constraints. Our contributions are threefold. First, we introduce a preparation scheme that recursively uses the Singular Value Decomposition (SVD) to construct the circuit from the MPS representation. This approach systematically parallelizes the disentangling operations, reducing the circuit depth from the conventional $O(n)$ to $O(\log n)$. Secondly, our method is designed for practical implementation; we provide a common example of an explicit mapping from the circuit to the topology of real quantum hardware, paving the way for generalizing the method to arbitrary chip topologies, thereby reducing the overhead of quantum circuit depth. Third, we address the primary source of error—two-qubit gates—by incorporating Cartan's KAK decomposition~\cite{vidal2003universal, tucci2005introduction}. This optimization reduces the required CNOT gates for the necessary two-qubit unitary transformations from three to two, achieving a one-third reduction in the two-qubit gate count for states with complex-valued amplitudes.
	
	To establish the theoretical foundation of our approach, we found a broad class of quantum states—those with an MPS representation of bounded bond dimension—for which our method guarantees efficient preparation. This result confirms the broad applicability and scalability of our framework for large-scale quantum systems. To validate our contributions, we perform extensive numerical experiments comparing our IMPS framework with existing MPS-based techniques. The results demonstrate substantial improvements in both circuit depth and scalability, highlighting the practical benefits of our approach for real-world quantum computing applications. Collectively, these developments establish our framework as a robust and versatile solution for quantum state preparation, adaptable to diverse hardware platforms and problem scales.
	
	In this paper, the term \emph{U-depth} denotes the number of sequential layers of two-qubit unitary gates that cannot be executed in parallel; this metric is directly proportional to the total circuit depth. The term \emph{layer}, in contrast, refers to the number of applications of the core preparation block within the overall MPS algorithm.
	
\section{Methods}
In this section, we outline our proposed method. We begin by reviewing standard MPS preparation techniques to provide the necessary background. Next, we derive a new representation for unitary operations acting on the quantum state. This allows us to create a method for disentangling any pair of qubits, making our approach compatible with various quantum chip layouts. We then describe how to implement the required two-qubit gates efficiently, rigorously proving that such gates can be realized using only two CNOT gates and several single-qubit gates. Finally, we identify a broad class of functions with bounded MPS rank, which can efficiently approximate nearly all smooth functions. We conduct numerical experiments with common probability distributions in the next chapter to validate our approach.
	
\subsection{Improved MPS and Circuit Depth Reduction}
A canonical MPS consisting of $n$ qubits with its state $\ket{\psi}$ can be written as:
\begin{align}
	\ket{\psi}&=\sum\limits_{\sigma_1,\cdots ,\sigma_n\in\{0,1\} }{c_{\sigma_1,\cdots ,\sigma_n}\ket{\sigma_1,\cdots ,\sigma_n}}\nonumber\\
	 &\approx \sum\limits_{a_{1}\cdots a_{n-1}}{\sum\limits_{s_{1}\cdots s_{n}}{A_{s_{1},a_{1}}^{[1]}A_{s_{2},a_{1}a_{2}}^{[2]}}}\cdots A_{s_{n},a_{n-1}}^{[n]}{\ket{s_{1},s_{2},\cdots ,s_{n}}},\nonumber
\end{align}
with the physical indices $\{s_{n}=0,1,\cdots,d-1\}$ and virtual indices $\{a_{n}=0,1,\cdots,\chi-1\}$, which are used for tensor contraction and joint computation. The virtual dimensions are typically bounded as $\dim(a_{n})\leq \chi$ to control computational cost. Relevant theories show that as long as $\chi$ is sufficiently large, it can represent any quantum state~\cite{rudolph2023decomposition}.

\begin{figure}[htbp]
		\centering  
		\begin{tikzpicture}
			\node[scale=0.9] {
				\begin{quantikz}[row sep=0.2cm]
					\centering
					\lstick{}&\gate[2][1cm]{U_{1}}&&&&&\\
					\lstick{}&&\gate[2][1cm]{U_{2}}&&&&\\
					\lstick{}&&&\gate[2][1cm]{U_{3}}&&&\\
					\lstick{}&&&&\gate[2][1cm]{U_{4}}&&\\
					\lstick{}&&&&&\gate[2][1cm]{U_{5}}&\\
					\lstick{}&&&&&&
				\end{quantikz}
			};
		\end{tikzpicture}
		\caption{Canonical MPS applied to 6 qubits with 5 times of fourth-order unitary matrix}
		\label{Fig.canonicalmps}
	\end{figure}
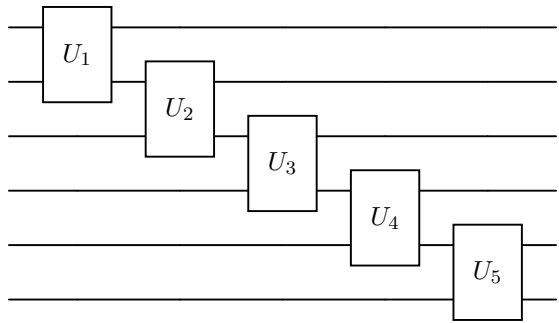

In this work, we restrict our attention to two-level quantum states and use only universal quantum gates on two qubits to realize entanglement and approximation. This constraint leads us to set $d=2$ and $\chi=2$. By performing singular value decomposition (SVD), we construct $n-1$ fourth-order unitary matrices and map them onto a quantum circuit that approximately disentangles $\ket{\psi}$ into $\ket{0}_n$. Then, by reversing this circuit, we obtain an approximate state preparation process. The canonical MPS quantum circuit diagram is shown in Fig.~\ref{Fig.canonicalmps}.

Similar to the conventional MPS approximation schemes, we aim to find a sequence of two-qubit unitary operations that approximately transform the quantum state $\ket{\psi}$ into the all-zero state $\ket{0}_n$. Once this transformation is accomplished, the inverse circuit serves as an approximate implementation of the unitary that prepares the target state, thus completing the approximate amplitude encoding.

To formalize our analysis, we introduce the notation $C_{a,b}(x, y)$ to represent a vector extracted from the amplitudes of the quantum state $\ket{\psi}$. Specifically, $C_{a,b}(x, y)$ denotes the subset of amplitudes in which the qubits at indices $a$ and $b$ are in states $x$ and $y$, respectively. The resulting vector has a length of $2^{n-2}$, where $n$ is the total number of qubits.

For instance, suppose $\ket{\psi}$ is a quantum state of a four-qubit system. In this case, the notation $C_{1,3}(1, 0)$ refers to the vector composed of amplitudes where the first and third qubits are in states $1$ and $0$, respectively:
\begin{align}
	C_{1,3}(1, 0)=[c_{0100},c_{0110},c_{1100},c_{1110}].\nonumber
\end{align}

Assume a two-qubit quantum gate acts on the qubits indexed by $a$ and $b$, which can be viewed as applying a $4\times4$ unitary matrix $U^{\prime}$ to the matrix $A_{a,b}$ constructed from the amplitudes of $\ket{\psi}$:
\begin{align}
	A_{a,b}=\left[\begin{matrix}
		C_{a,b}(0, 0)\\[0.5ex]
		C_{a,b}(0, 1)\\[0.5ex]
		C_{a,b}(1, 0)\\[0.5ex]
		C_{a,b}(1, 1)
	\end{matrix}\right].
	\label{eq4}
\end{align}
Here, $A_{a,b}$ is a $4 \times 2^{n-2}$ matrix.

By performing a singular value decomposition (SVD) on the matrix $A_{a,b}$ as $A_{a,b} = USV^{\dagger}$, and then applying the inverse unitary matrix $U^{-1}$ to eliminate the $U$ obtained from the decomposition, the resulting amplitude transformation becomes:
\begin{align}
	&U^{-1}A_{a,b}=U^{-1}USV^{\dagger}\\
	=&\left[\begin{matrix}
		\lambda_0&0&0&0\\[0.5ex]
		0&\lambda_1&0&0\\[0.5ex]
		0&0&\lambda_2&0\\[0.5ex]
		0&0&0&\lambda_3
	\end{matrix}\right]\left[\begin{matrix}
		V_{00}\\[0.5ex]
		V_{01}\\[0.5ex]
		V_{10}\\[0.5ex]
		V_{11}
	\end{matrix}\right]=\left[\begin{matrix}
		\lambda_0V_{00}\\[0.5ex]
		\lambda_1V_{01}\\[0.5ex]
		\lambda_2V_{10}\\[0.5ex]
		\lambda_3V_{11}
	\end{matrix}\right].
\end{align}
Here $\lambda_{0} \geq \lambda_{1} \geq \lambda_{2} \geq \lambda_{3} \geq 0$, and all vectors $V_{ij}$ are pairwise orthogonal and normalized. The squared norms of the first two row vectors sum to $\lambda_{0}^{2} + \lambda_{1}^{2}$, which corresponds to preserving the dominant components of the quantum state's squared amplitude. By truncating the last two rows, the state of qubit $a$ can be approximately set to $\ket{0}$. 

In particular, if the sum of the squared amplitudes in the first two rows equals 1, then complete disentanglement is achieved, as qubit $a$ will be exactly in the state $\ket{0}$. 

If we perform the disentangling sequentially according to the qubit order and normalize the resulting vectors at each step, the entire process becomes equivalent to the canonical MPS-based approximate amplitude encoding scheme. Since the qubit pair $(a,b)$ can be arbitrarily chosen, this method enables disentangling to be performed on any pair of qubits, allowing the implementation of two-qubit gates at arbitrary positions in the quantum circuit.

By leveraging this technique, we perform parallel disentangling to reduce the overall depth of the quantum circuit, thereby establishing the following proposition:

\begin{proposition}[IMPS]
Improved MPS (IMPS) achieve efficient amplitude preparation with $\mathcal{O}(\log n)$ circuit depth. When considering hardware connectivity constraints, the method supports approximate amplitude preparation on arbitrarily connected topological structures.
\end{proposition}

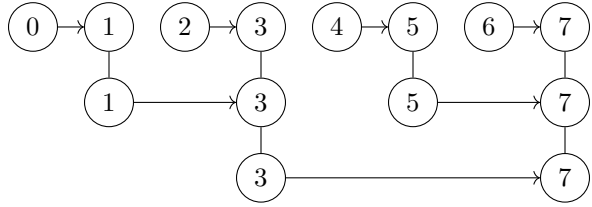
\begin{figure}[h]
	\centering
	\begin{tikzpicture}[scale=1, font=\small]
		\node[draw, circle] (0) at (0,3) {0};
		\node[draw, circle] (1) at (1,3) {1};
		\node[draw, circle] (2) at (2,3) {2};
		\node[draw, circle] (3) at (3,3) {3};
		\node[draw, circle] (4) at (4,3) {4};
		\node[draw, circle] (5) at (5,3) {5};
		\node[draw, circle] (6) at (6,3) {6};
		\node[draw, circle] (7) at (7,3) {7};
		\draw[->] (0) -- (1);
		\draw[->] (2) -- (3);
		\draw[->] (4) -- (5);
		\draw[->] (6) -- (7);
		\node[draw, circle] (11) at (1,2) {1};
		\node[draw, circle] (33) at (3,2) {3};
		\node[draw, circle] (55) at (5,2) {5};
		\node[draw, circle] (77) at (7,2) {7};
		\draw[-] (1) -- (11);
		\draw[-] (3) -- (33);
		\draw[-] (5) -- (55);
		\draw[-] (7) -- (77); 
		\draw[->] (11) -- (33);
		\draw[->] (55) -- (77);
		\node[draw, circle] (333) at (3,1) {3};
		\node[draw, circle] (777) at (7,1) {7};
		\draw[-] (33) -- (333);
		\draw[-] (77) -- (777); 
		\draw[->] (333) -- (777);
	\end{tikzpicture}
	\caption{Quantum circuit illustrating disentanglement with logarithmic depth, exemplified using 8 qubits. The circuit comprises three U-depth: the first disentangles qubits 0, 2, 4, and 6; the second disentangles qubits 1 and 5; and the third disentangles qubit 3. This process effectively approximates the overall amplitude distribution toward the $\ket{0}_3$ state.}
	\label{fig:8exampletop}
\end{figure}

First, let us consider the scenario without hardware topological constraints. On a system of $n$ qubits, we can apply $\lfloor n/2\rfloor$ two-qubit gates in parallel in each layer, disentangling $\lfloor n/2\rfloor$ qubits at each step. By applying this procedure iteratively, all qubits can be approximately disentangled with a circuit depth of only $\mathcal{O}(\log n)$. The reverse of this disentanglement process can be viewed as the approximate preparation of the target quantum state. This disentanglement structure is known as a Tree Tensor Network (TTN), as illustrated in Fig.~\ref{fig:8exampletop}.

\begin{figure}[h]
	\centering
	\begin{tikzpicture}[scale=3] % 调整缩放比例
		% 定义正方体的8个顶点坐标（x,y,z）
		\coordinate (A) at (0,0,0); % 底面前左下
		\coordinate (B) at (1,0,0); % 底面前右下
		\coordinate (C) at (1,1,0); % 后面右下
		\coordinate (D) at (0,1,0); % 后面左下
		\coordinate (E) at (0,0,1); % 前面左上
		\coordinate (F) at (1,0,1); % 前面右上
		\coordinate (G) at (1,1,1); % 后面右上
		\coordinate (H) at (0,1,1); % 后面左上
		
		\draw[thick,blue,->] (G) -- (H);
		\draw[thick,blue,->] (F) -- (E);
		\draw[thick,blue,->] (C) -- (D);
		\draw[thick,blue,->] (B) -- (A);
		\draw[thick,green,->] (H) -- (E);
		\draw[thick,green,->] (G) -- (F);
		\draw[thick,green,->] (D) -- (A);
		\draw[thick,green,->] (C) -- (B);
		\draw[thick,red,->] (E) -- (A);
		\draw[thick,red,->] (F) -- (B);
		\draw[thick,red,->] (G) -- (C);
		\draw[thick,red,->] (H) -- (D);
		% 绘制空心节点并标注
		\foreach \point/\label in {A/$q_{000}$, B/$q_{100}$, C/$q_{110}$, D/$q_{010}$, E/$q_{001}$, F/$q_{101}$, G/$q_{111}$, H/$q_{011}$} {
			\fill[thick] (\point) circle (0.5pt); % 绘制空心圆节点
			\node[shift={(0.6,0.2,0)}] at (\point) {\label}; % 标注节点名称
		}
	\end{tikzpicture}
	\caption{Illustration of an 8-qubit HTN circuit connectivity scheme for disentanglement, where each color denotes a group of parallel two-qubit gates. The scheme is constructed by sorting qubit indices in binary order and mapping each binary digit from “1” to “0.” For qubit counts that are not powers of two, the corresponding vertices and their connecting arrows are omitted accordingly.}
	\label{fig:sexampletop}
\end{figure}
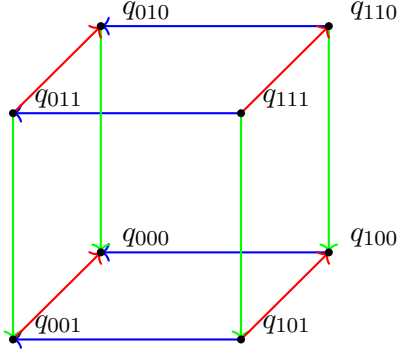

In this study, we propose a disentangling approach that is denser than the TTN. This method, under specific chip topologies, does not incur additional circuit depth compared to the TTN framework; we designate this structure as the Hypercube Tensor Network (HTN). Numerical simulations indicate that the proposed method exhibits superior fidelity. To elucidate the HTN connectivity scheme, we reorder the qubits using binary indexing, mapping them to the vertices of a hypercube. A specific example is provided in Fig.~\ref{fig:sexampletop}.

When considering a realistic chip topology, we model it as a connected graph. We then iteratively contract the edges of this graph, systematically reducing it to a single vertex. This contraction process corresponds to a disentanglement scheme using two-qubit gates. The principles of graph theory ensure that non-interfering operations (i.e., those on non-adjacent edges) can be executed in parallel within the quantum circuit, thereby enhancing disentanglement efficiency. The specific contraction scheme can be determined using computational methods, such as reinforcement learning or optimization algorithms, to find the optimal sequence of edge contractions~\cite{ostaszewski2021reinforcement,fosel2021quantum,fosel2021quantum}.

\subsection{Two-Qubit Gate Optimization}
In this section, we investigate the role of universal two-qubit gates in the disentanglement process within the IMPS framework, thereby deriving a class of equivalent matrix representations. Within this class, we identify a unitary scheme that can be implemented using only two CNOT gates, as opposed to the three CNOT gates typically required. This approach effectively reduces the circuit depth for the corresponding unitary matrix. Given that implementation challenges on physical quantum chips predominantly stem from two-qubit gates, our optimization strategy significantly mitigates the complexity of loading complex-valued amplitudes. This is because the SVD of real-valued amplitudes yields orthogonal matrices, which can already be implemented efficiently with two CNOT gates and single-qubit gates, thus requiring no such optimization.

	First we consider following operation:
	\begin{align}
		\label{equ1u2}
		\begin{bmatrix}
			U_{1} & O \\[0.5ex]
			O & U_{2}
		\end{bmatrix}
		\begin{bmatrix}
			\lambda_0V_{00}\\[0.5ex]
			\lambda_1V_{01}\\[0.5ex]
			\lambda_2V_{10}\\[0.5ex]
			\lambda_3V_{11}
		\end{bmatrix}.
	\end{align}
	
	Following the application of this matrix as shown in~\ref{equ1u2}, and due to the unitarity of $U_1$, the sum of the squared moduli of the amplitudes in the first two rows remains maximal. This enables the identification of an equivalence class of matrices of the form 
	$
	\begin{bmatrix} U_1 & O \\[0.5ex] O & U_2 \end{bmatrix}U^{-1}, 
	$
	which can replace $U^{-1}$ in the amplitude encoding process, where $U_1$ and $U_2$ are arbitrary second-order unitary matrices. The objective is to identify, within this equivalence class, the fourth-order unitary matrix that can be implemented most efficiently using quantum circuits.
	
	We begin by considering a slightly modified version of the cosine-sine decomposition of $U^{-1}$, which yields:
	\begin{align}
		U^{-1}=
		\begin{bmatrix}
			A^{\prime} & O \\[0.5ex]
			O & B^{\prime}
		\end{bmatrix}\begin{bmatrix}
			C & S \\[0.5ex]
			S & -C
		\end{bmatrix}\begin{bmatrix}
			A & O \\[0.5ex]
			O & B
		\end{bmatrix}.
	\end{align}
	Here, $A^{\prime}$, $B^{\prime}$, $A$, and $B$ are second-order unitary matrices, while $C$ and $S$ are real, diagonal, positive semi-definite matrices of order two that satisfy the condition $C^2 + S^2 = I$.
	
	Given the arbitrariness of matrices $U_1$ and $U_2$, we may set $U_1 = A^{-1}A^{\prime -1}$ and $U_2 = B^{-1}B^{\prime -1}$, resulting in the matrix \begin{align}
		U_{2\mathrm{cx}}=
		\begin{bmatrix}
			A^{-1} & O \\[0.5ex]
			O & B^{-1}
		\end{bmatrix}\begin{bmatrix}
			C & S \\[0.5ex]
			S & -C
		\end{bmatrix}\begin{bmatrix}
			A & O \\[0.5ex]
			O & B
		\end{bmatrix},\label{u2cx}
	\end{align} 
	which can be efficiently implemented in a quantum circuit. We will demonstrate that this unitary matrix can be realized using two CNOT gates and a few single-qubit gates. The proof is provided in Appendix~\ref{appendix-proof}.
	
	We present the numerical test results for this case in Section~\ref{section3}, where a comprehensive comparison is conducted. Furthermore, we significantly reduce the two-qubit gate depth required for MPS amplitude preparation. This simplification is formally articulated in Theorem~\ref{theorem2} below.

\begin{theorem}[Fourth-Order Unitary Decomposition Theorem]Any fourth-order unitary matrix arising from the SVD in MPS amplitude preparation can be decomposed into two CNOT gates and a sequence of single-qubit unitaries. 
	\label{theorem2}
\end{theorem}

\subsection{Amplitudes with Bounded MPS Rank}
In this section, we identify a class of functions that can be represented by states with a bounded Matrix Product State (MPS) rank. An MPS represents the wavefunction $|\psi\rangle = \sum_{\sigma_1, \dots, \sigma_n} c_{\sigma_1 \dots \sigma_n} |\sigma_1 \dots \sigma_n\rangle$, with coefficients given by: 
\[		\vspace{-0.5em}
c_{\sigma_1 \dots \sigma_n} = A_{\sigma_1}^{[1]} A_{\sigma_2}^{[2]} \cdots A_{\sigma_n}^{[n]},		\]
where $A_{\sigma_i}^{[i]} \in \mathbb{C}^{\chi_{i-1} \times \chi_i}$ for $i = 2, \dots, n-1$, $A_{\sigma_1}^{[1]} \in \mathbb{C}^{1 \times \chi_1}$, and $A_{\sigma_n}^{[n]} \in \mathbb{C}^{\chi_{n-1} \times 1}$. The bond dimension at site $i$, denoted by $\chi_i$, is the dimension of the index connecting sites $i$ and $i+1$. The MPS rank of the state is the maximum bond dimension, $\chi = \max_i \chi_i$, which reflects the maximum entanglement across any bipartition and is determined by the Schmidt rank:
\vspace{-0.5em}
\[
|\psi\rangle = \sum_{\alpha=1}^{\chi_i} s_\alpha |\phi_\alpha^L\rangle |\phi_\alpha^R\rangle.
\vspace{-0.5em}\]

Neglecting the normalization factor required for quantum states, we can establish relations for the MPS rank under element-wise addition and multiplication. These are formalized in the following well-known theorem for tensor networks:

\begin{theorem}[MPS Rank over a Ring Structure]\label{theorem3}
	Neglecting normalization factors, we can treat the element-wise addition and multiplication of quantum states as operations within a ring structure. Let $\chi_f$ denote the MPS rank of a state corresponding to a function $f$ evaluated at $2^n$ uniformly discretized points. Then, the following subadditivity and submultiplicativity properties hold:
	\vspace{-0.5em}
	\begin{align}
		\chi_{f\pm g} \leq \chi_{f} + \chi_{g},\nonumber\\
		\chi_{f\times g} \leq \chi_{f} \times \chi_{g}.\nonumber
	\end{align}
\end{theorem}
	
As an additional outcome of the IMPS scheme, we can precisely prepare states with an MPS rank of 2. This includes states representing the cosine function $a\cos(bx) + c$, the linear function $ax + b$, the $n$-qubit GHZ state, and the W-state, all with a circuit depth of $\mathcal{O}(\log n)$. It can also be noted that a function like $f(x) = a e^{bx}$ has an MPS rank of 1 and can be implemented using only a single layer of parallel RY gates~\cite{zhuang2023quantum}. We consider the ring structure generated by these exponential, cosine, and linear functions, which is significant as it encompasses many common smooth functions, making their efficient preparation highly valuable. Notably, functions generated from this ring structure are naturally related to Laplace, Taylor, and Fourier expansions and also serve as solutions to linear differential equations with constant coefficients.

Our subsequent experiments are conducted on functions generated from this ring structure. General continuous functions can often be effectively represented by the leading terms of their Fourier or Taylor expansions, which implies their MPS rank can be considered approximately bounded. These properties enable the efficient preparation of such functions using our method, a capability that is tested and validated in the numerical experiments in Section~\ref{section3}.

\section{Numerical Experiments}\label{section3}
	In this section, we consider the following three representative functions with bounded MPS rank:
	\begin{align}
		f_{1}(x)&=x(\mathrm{e}^{0.68x}+\mathrm{e}^{-2x}-0.7)\sin(24x),\nonumber \\
		f_{2}(x)&=(x^{2}-0.8x+0.04)\mathrm{e}^{-1.3x}\cos(7.2x - 1.6),\nonumber\\
		f_{3}(x)&=(x+\sin(13x)+\mathrm{e}^{-6.4x})\sin(2.8x + 14.3).\nonumber
	\end{align}
	We also include the Gaussian, Log-Normal, and Cauchy distributions as benchmarks for comparison:
	\[
	\begin{aligned}
		g_{1}(x)&=\frac{1}{\sqrt{2\pi}}\mathrm{e}^{-\frac{x^2}{2}}, \nonumber\\
		g_{2}(x)&=\frac{1}{x\sqrt{2\pi}}\mathrm{e}^{-\frac{\log^2(x)}{2}},\nonumber\\
		g_{3}(x)&=\frac{1}{\pi}\frac{1}{x^2+1}.\nonumber
	\end{aligned}
	\]
	The infidelity between pure quantum states is defined as:
	\begin{align}
		1 - \mathcal{F}(\ket{\phi}, \ket{\psi}) = 1 - |\langle \phi | \psi \rangle|^2 . \nonumber
	\end{align}

	In this paper, we compare circuits for amplitude preparation using MPS, TTN, and HTN. To ensure a fair comparison of fidelity at the same circuit depth, we also employ a Hardware-Efficient Network (HEN) scheme, which is an improved approach for chain-like chip architectures. This method fills the idle time slots in the traditional MPS circuit where no quantum gates are applied, thereby matching the circuit depth of the MPS scheme. It is important to clarify that the truncation methods for the MPS and TTN schemes differ from those for HTN and HEN. For MPS and TTN, truncation involves resetting the amplitude to its original length after each layer. In contrast, for HTN and HEN, the amplitude length is reset after each U-depth. This is an unavoidable difference arising from the distinct graph structures associated with the two classes of disentanglement methods.
	
\begin{figure}[htbp]
		\centering  
		\includegraphics[width=0.95\textwidth]{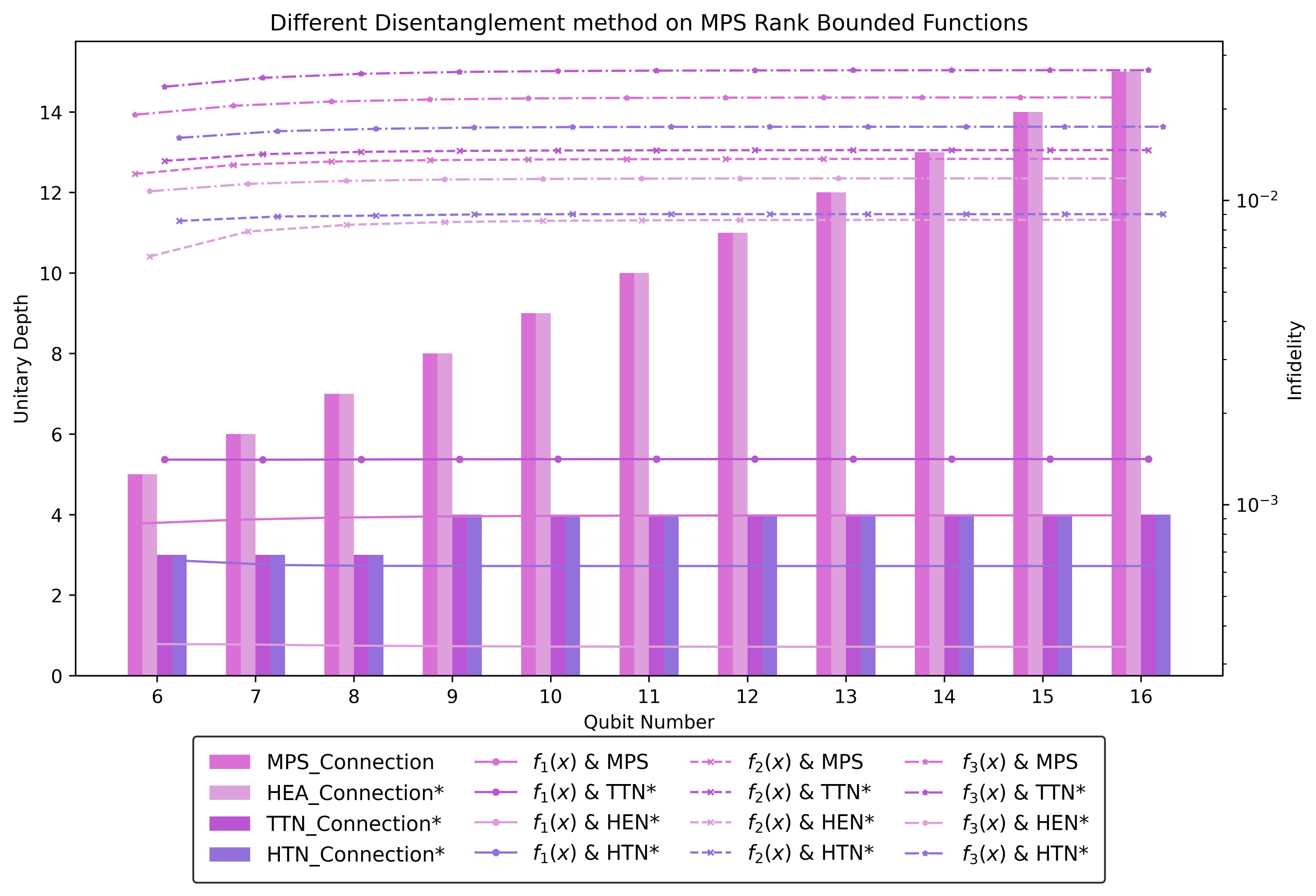}
		\vspace{2em}
		\includegraphics[width=0.95\textwidth]{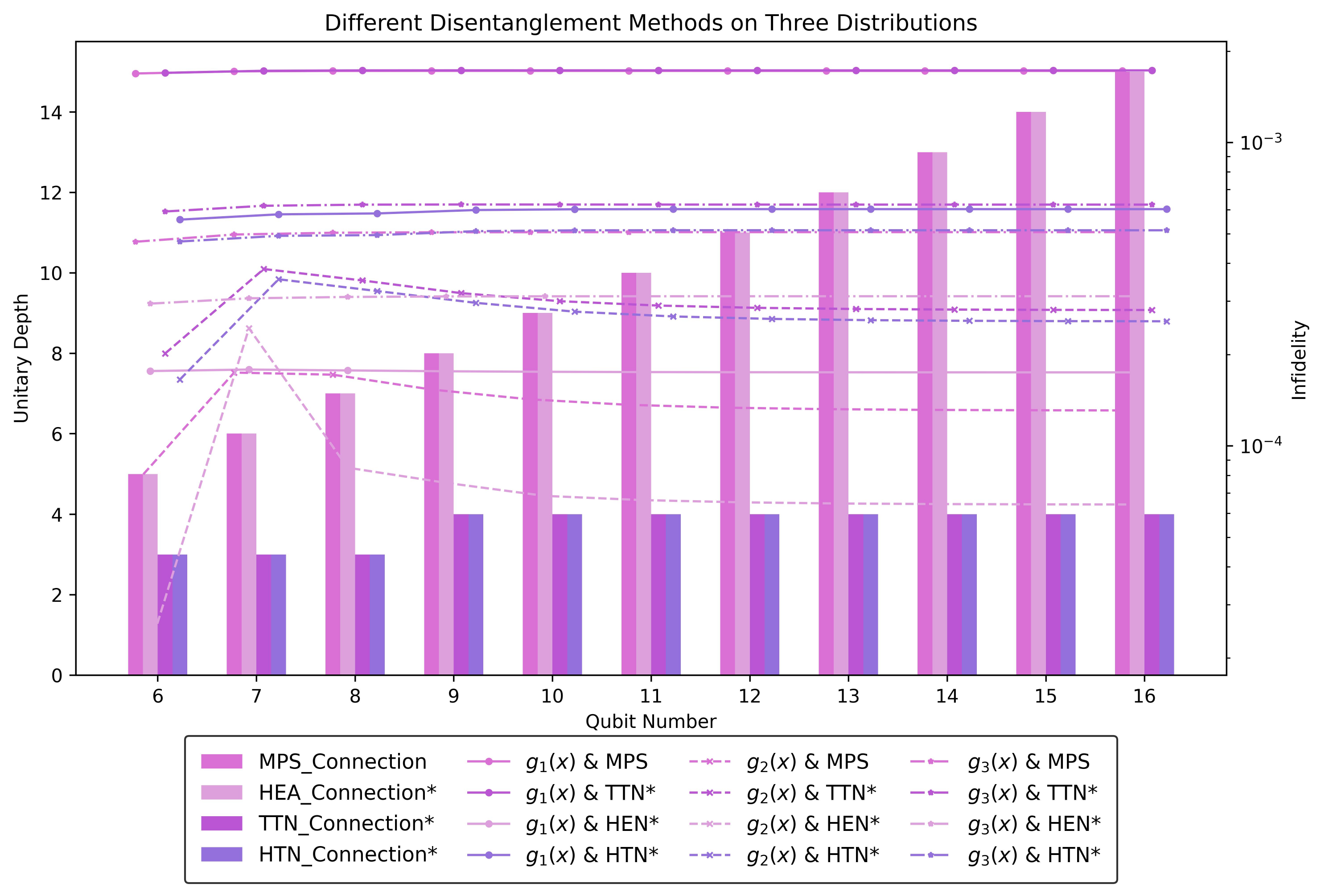}
		\caption{Infidelity of quantum state amplitude preparation for the functions $f_{1}(x)$, $f_{2}(x)$, $f_{3}(x)$, $g_{1}(x)$, $g_{2}(x)$, and $g_{3}(x)$ as the number of qubits increases. Among the four quantum state preparation schemes evaluated, the HTN method demonstrates significant advantages in terms of fidelity and circuit depth.}
		\label{stable}
	\end{figure}

Subsequent results demonstrate that our methods achieve higher fidelity. In all results, an asterisk (*) is used to highlight the primary methods developed in this study. We first present the specific numerical results for the six functions and distributions in Fig.~\ref{stable}.

\begin{figure}[htbp]
		\captionsetup{skip=4pt}
		\centering  
		\includegraphics[width=0.23\textwidth]{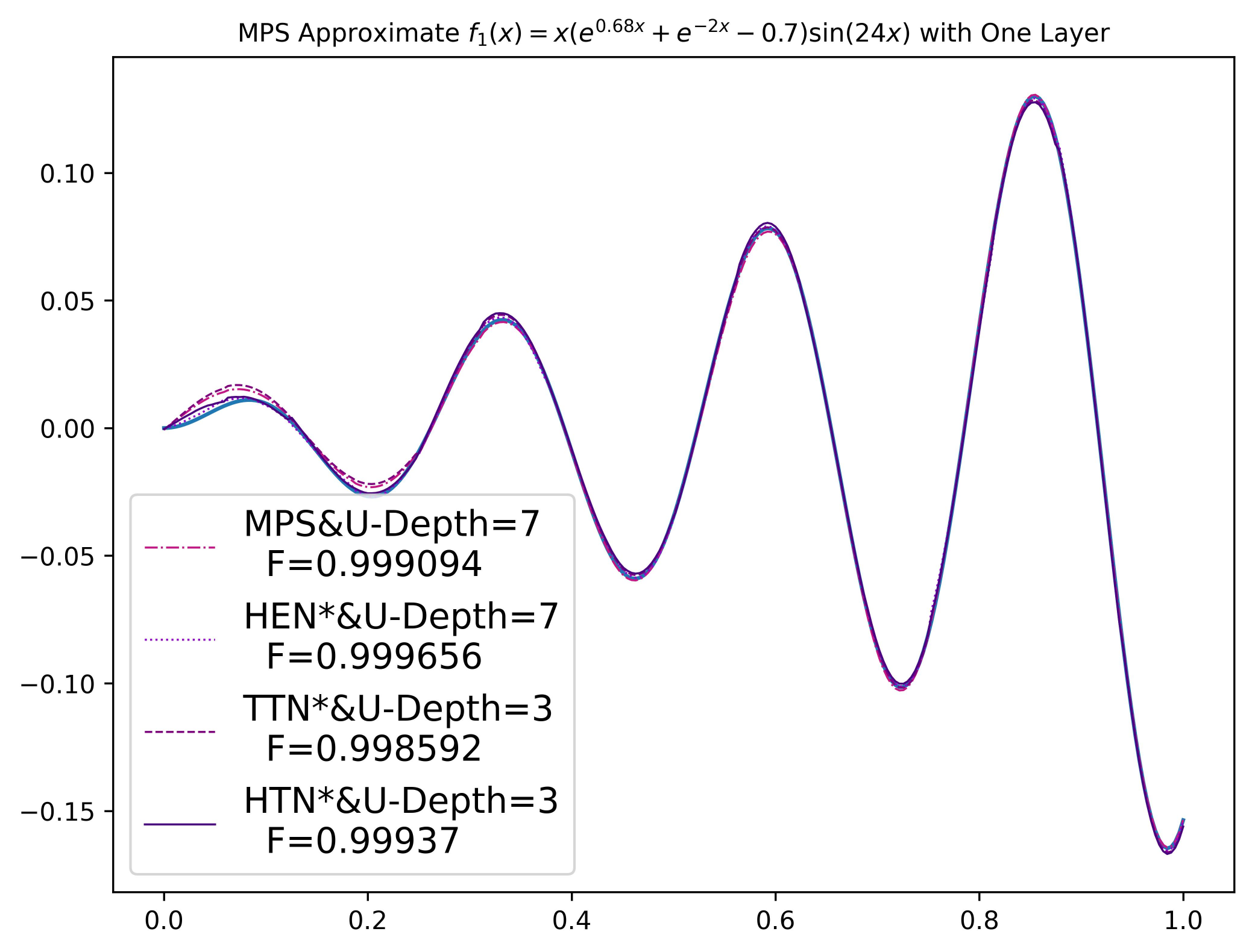}
		\includegraphics[width=0.23\textwidth]{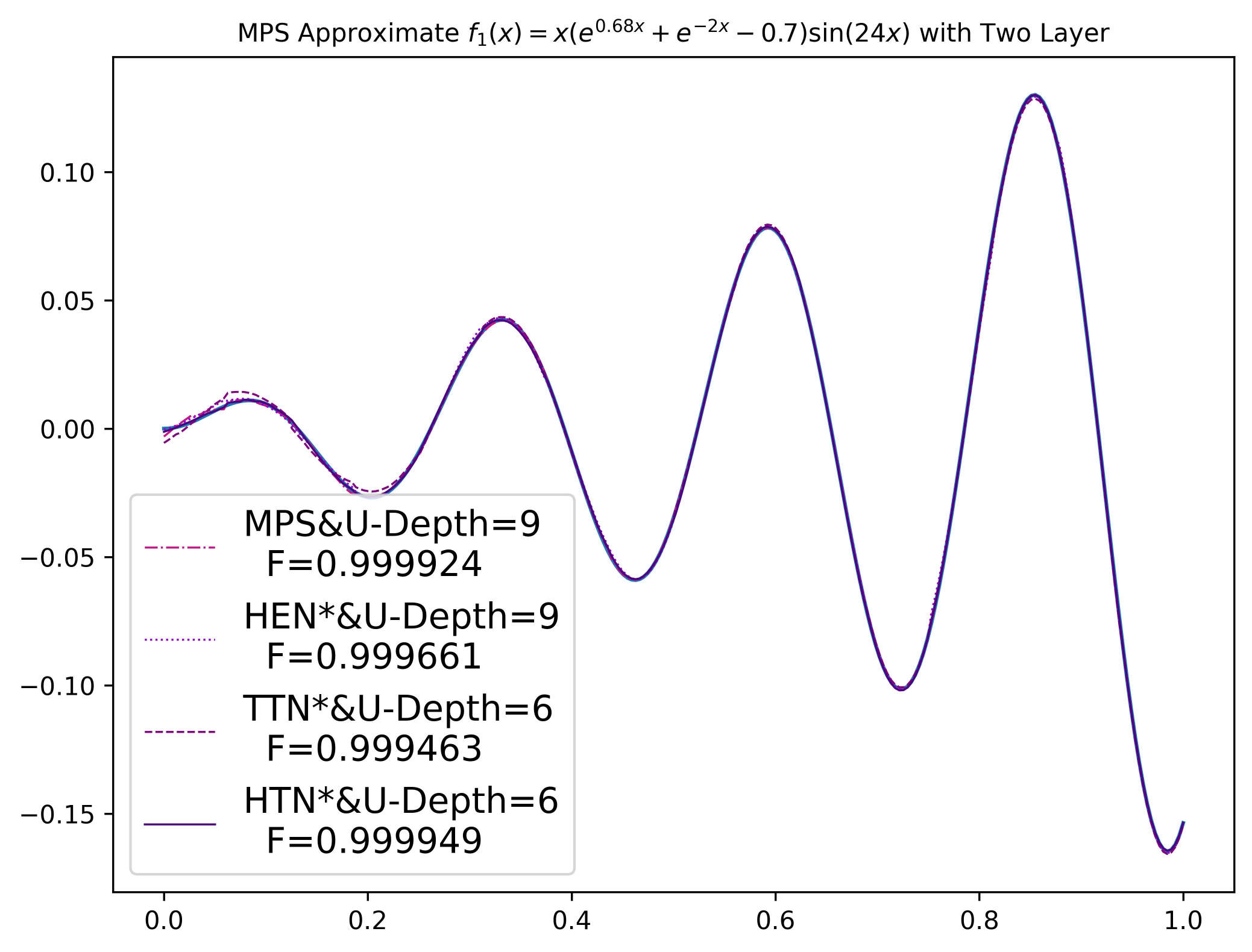}
		\includegraphics[width=0.23\textwidth]{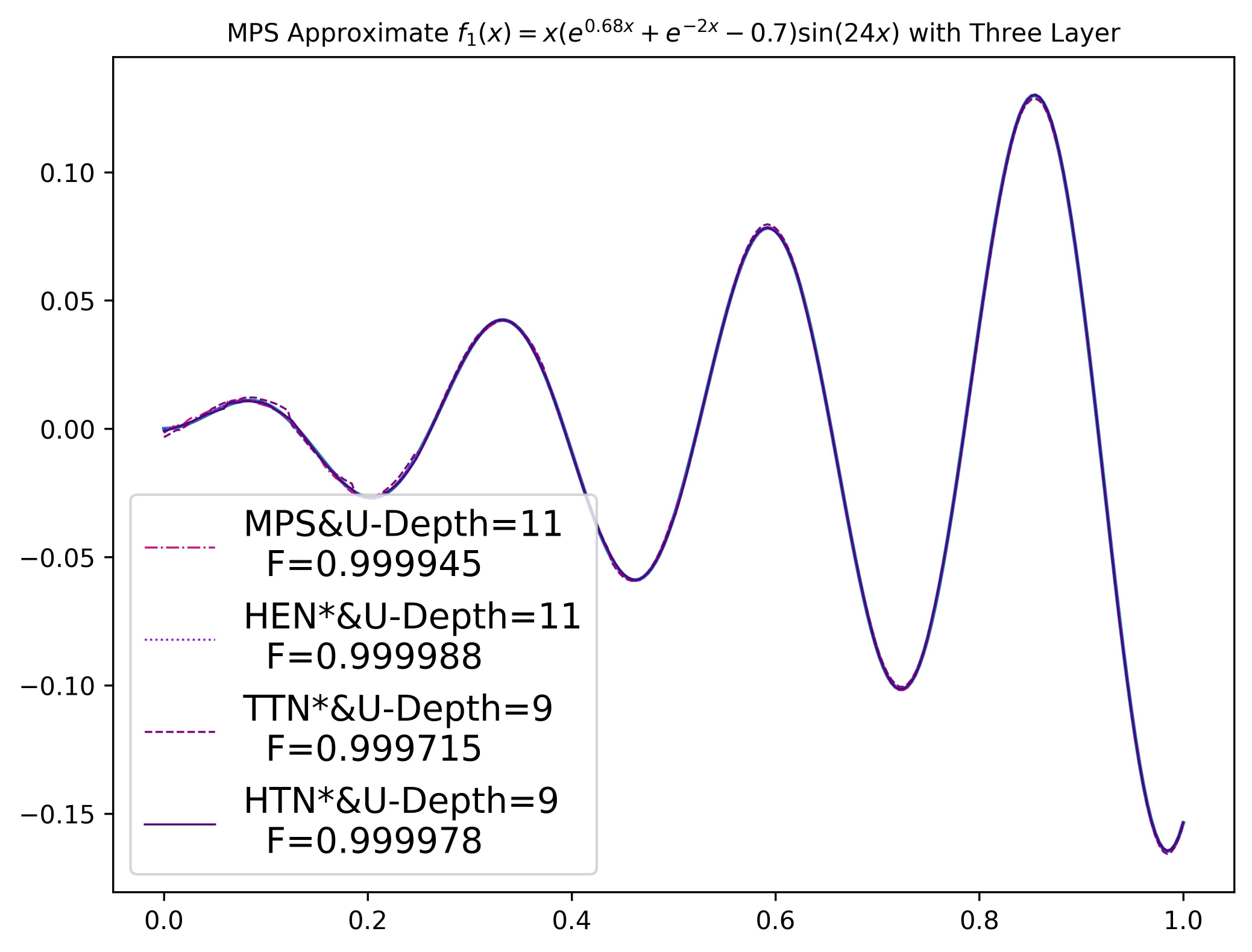}
		\includegraphics[width=0.23\textwidth]{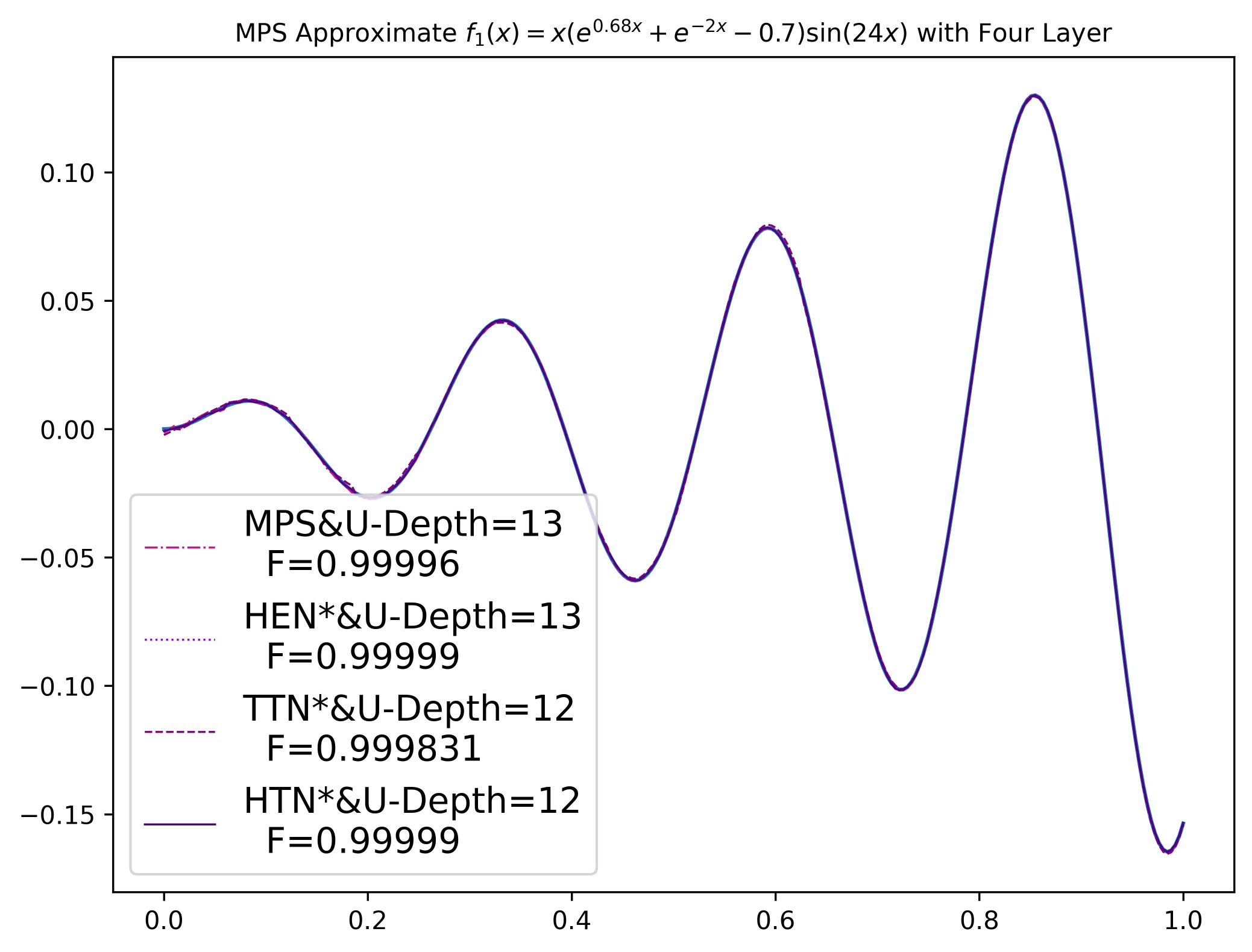}\\[1ex]
		\includegraphics[width=0.23\textwidth]{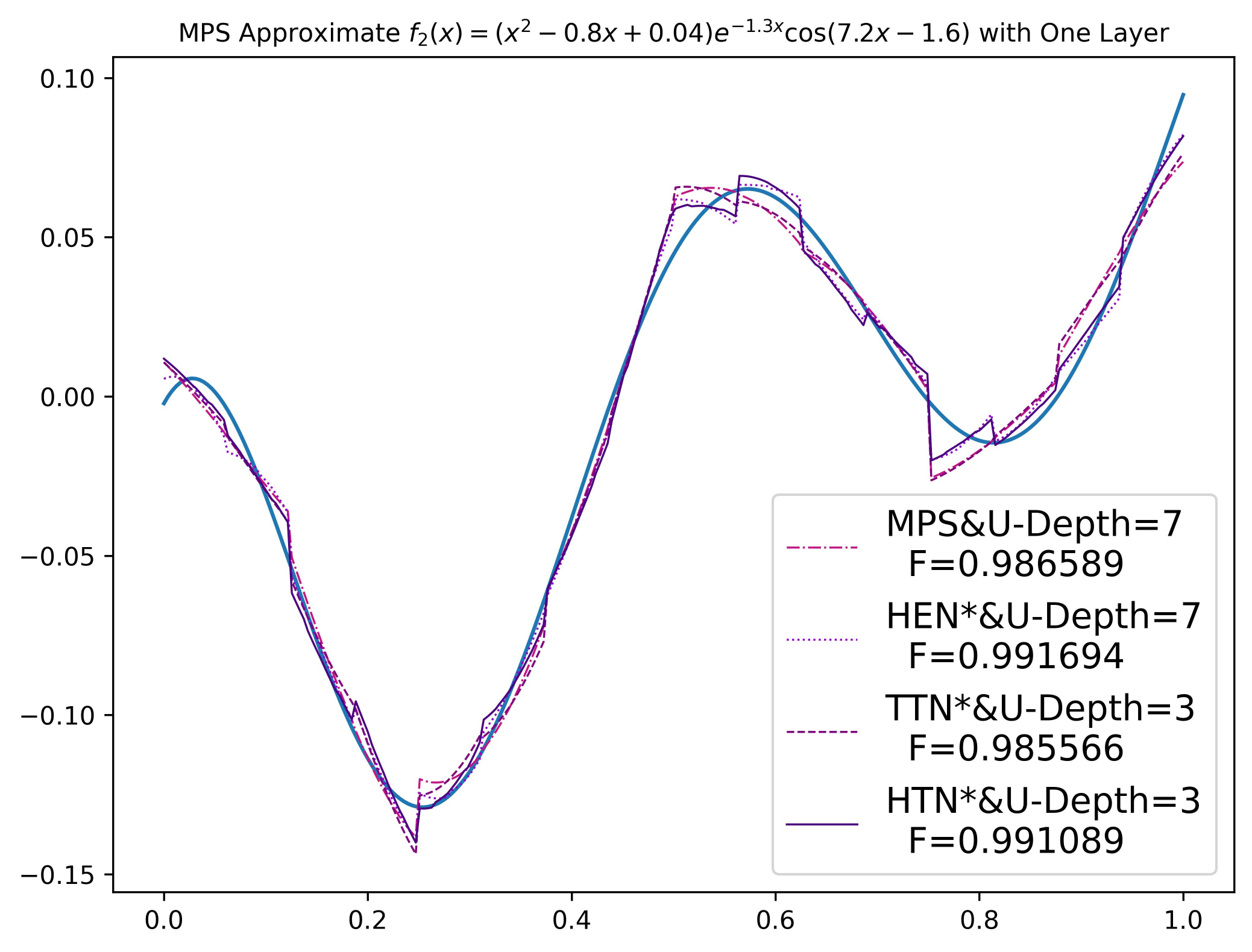}
		\includegraphics[width=0.23\textwidth]{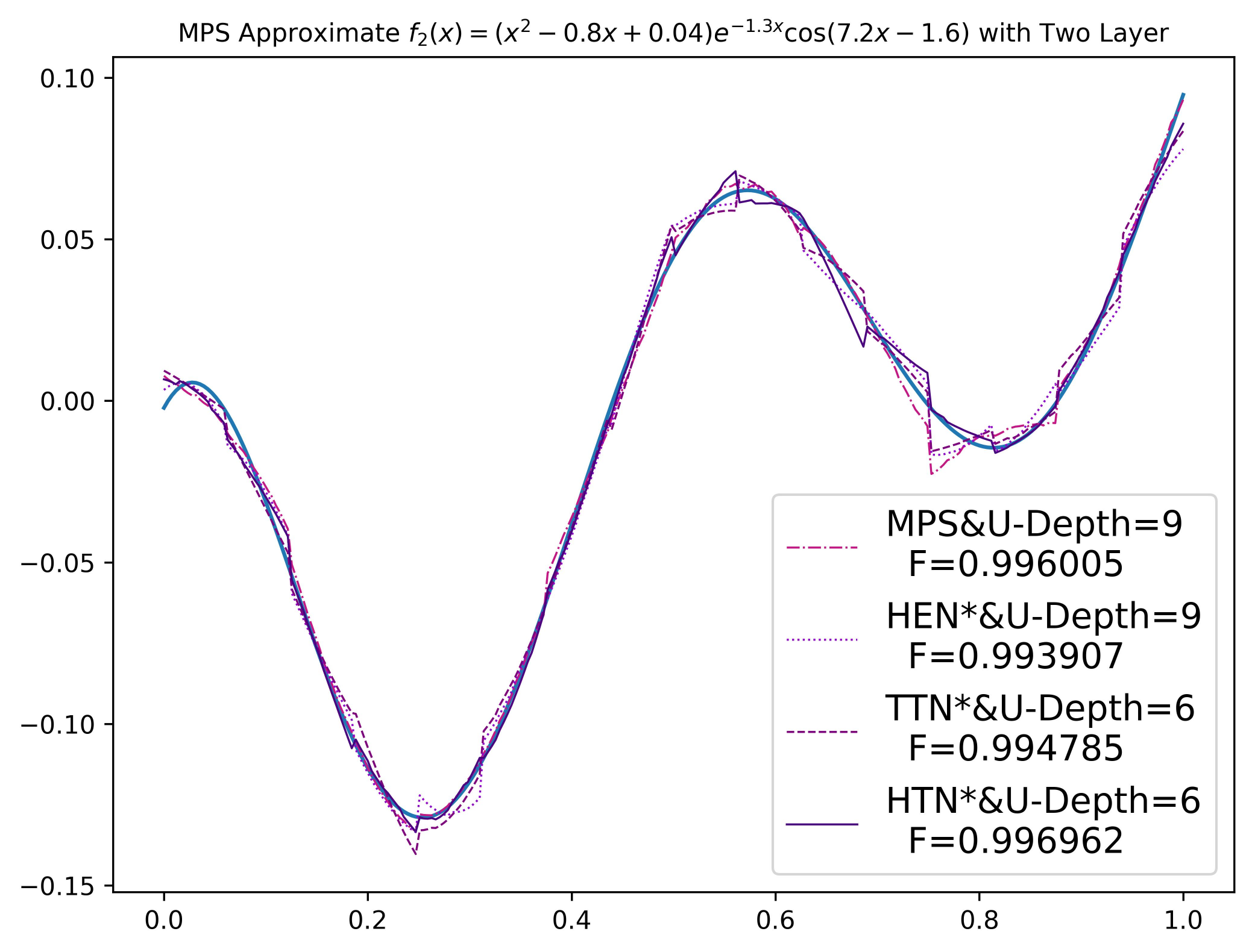}
		\includegraphics[width=0.23\textwidth]{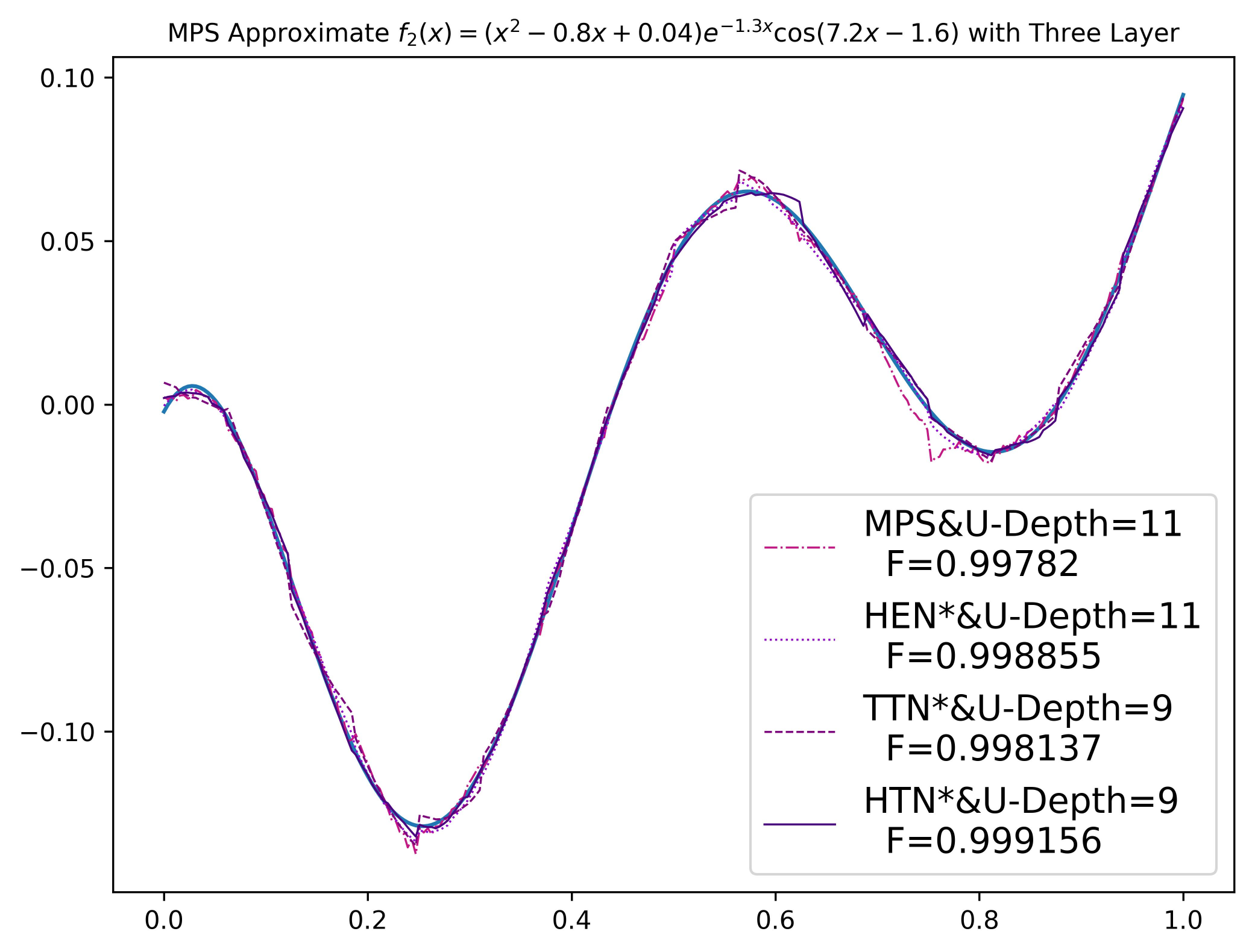}
		\includegraphics[width=0.23\textwidth]{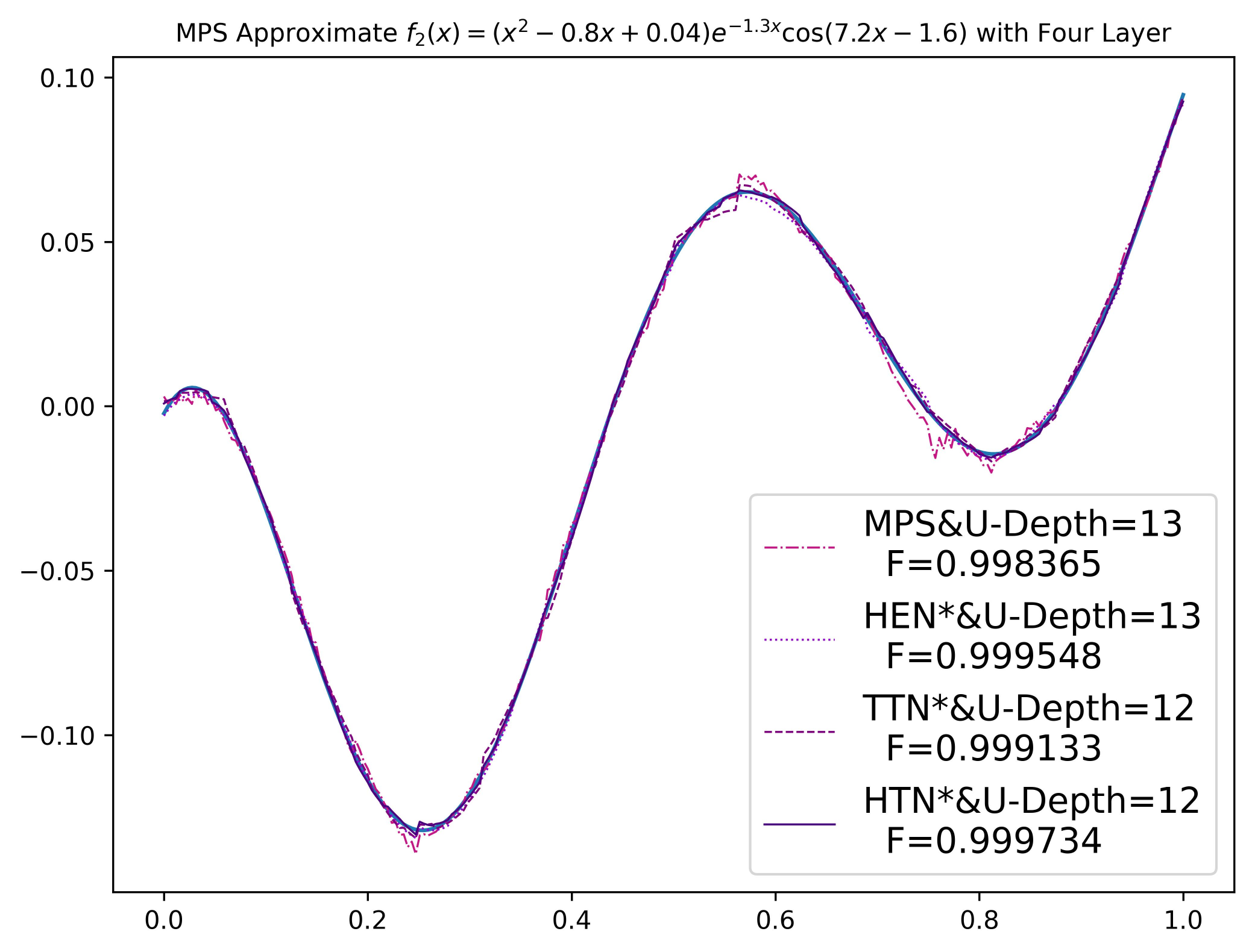}\\[1ex]
		\includegraphics[width=0.23\textwidth]{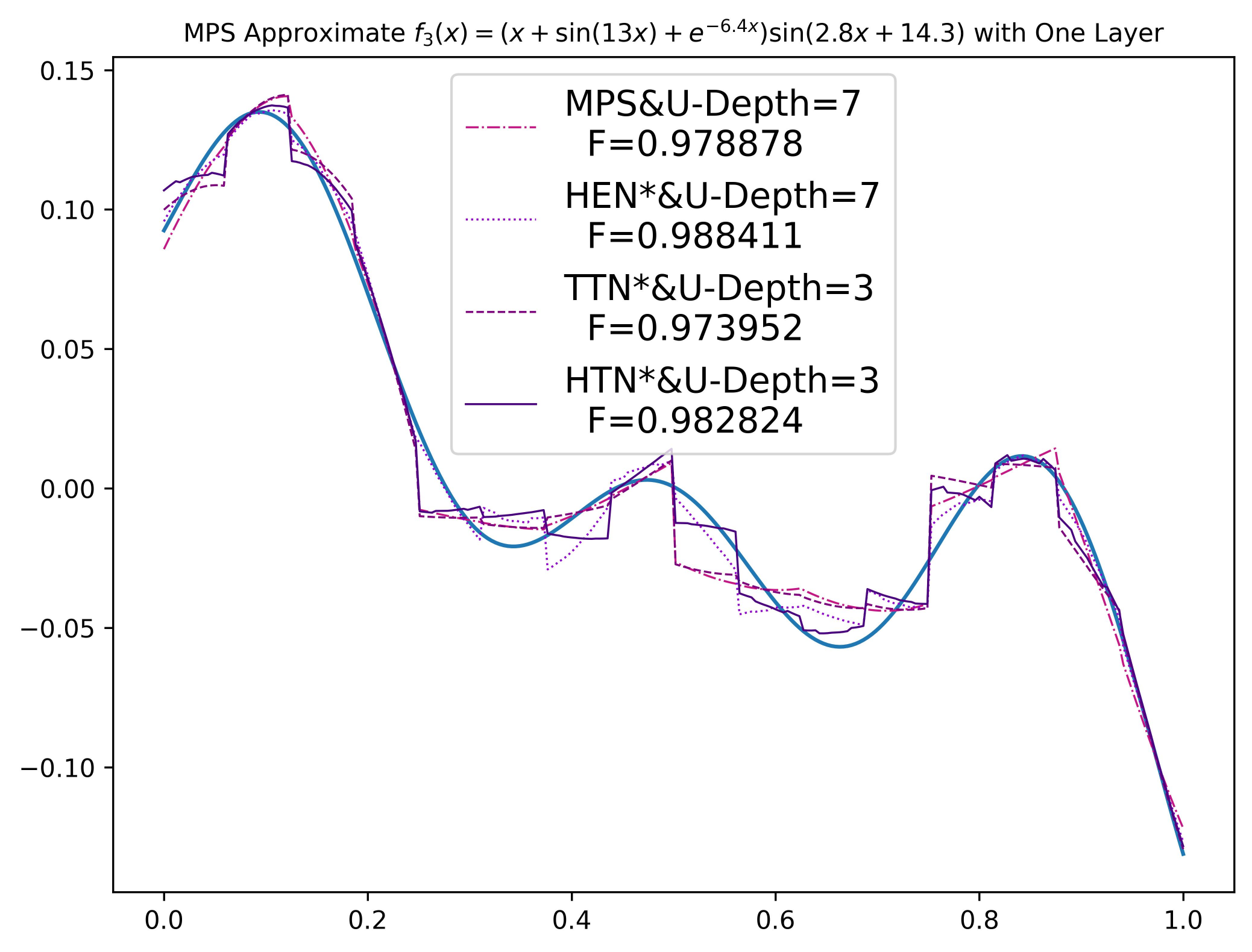}
		\includegraphics[width=0.23\textwidth]{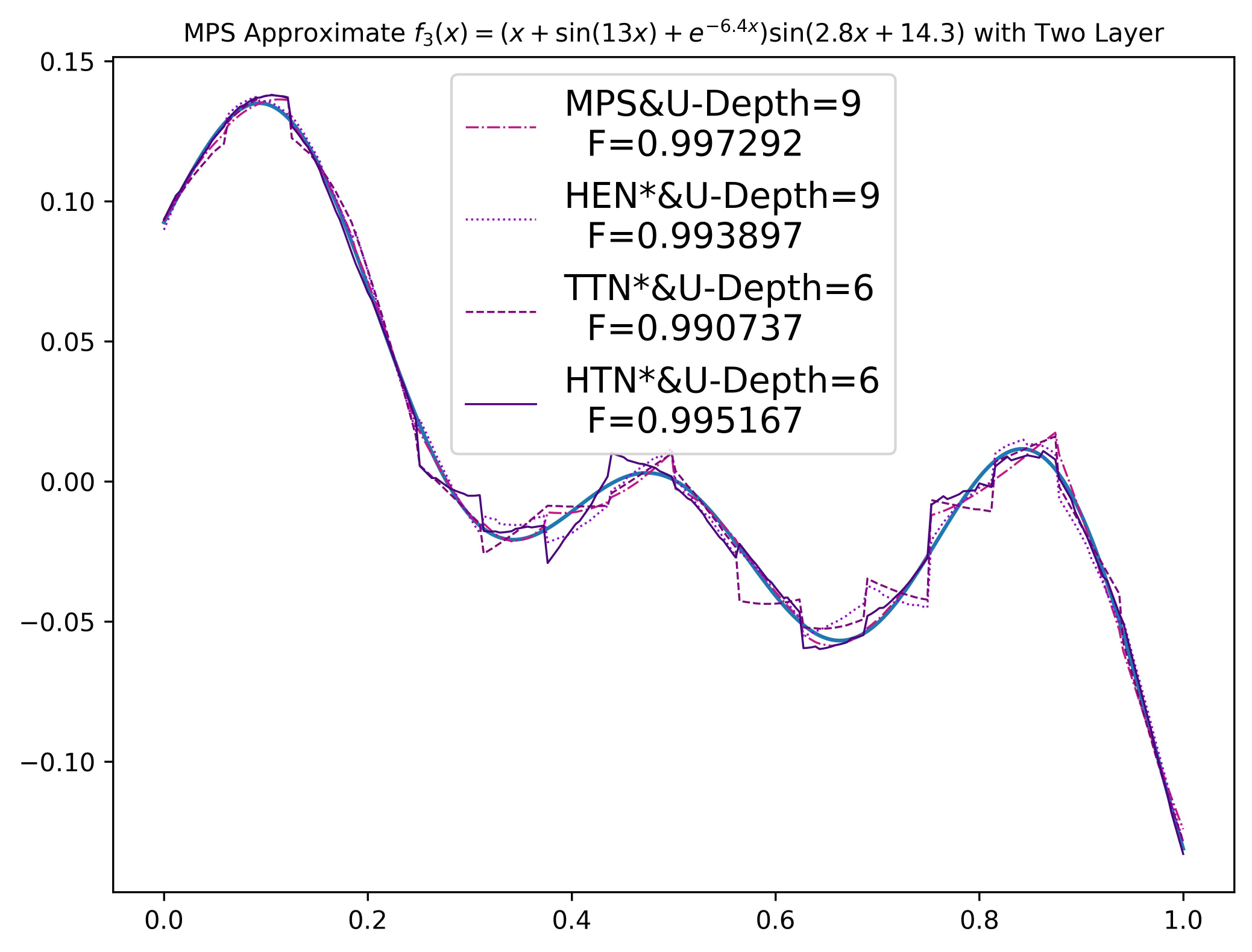}
		\includegraphics[width=0.23\textwidth]{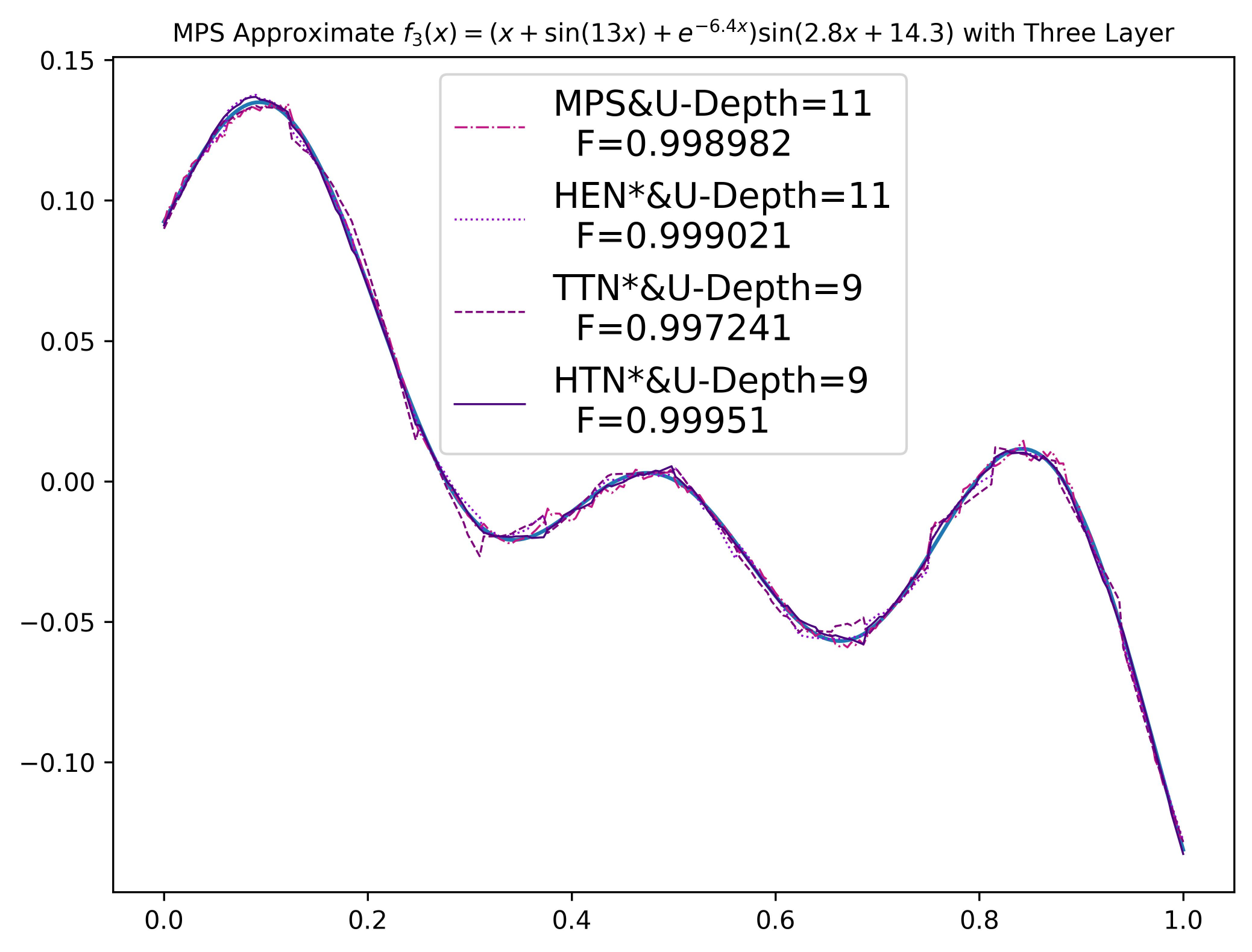}
		\includegraphics[width=0.23\textwidth]{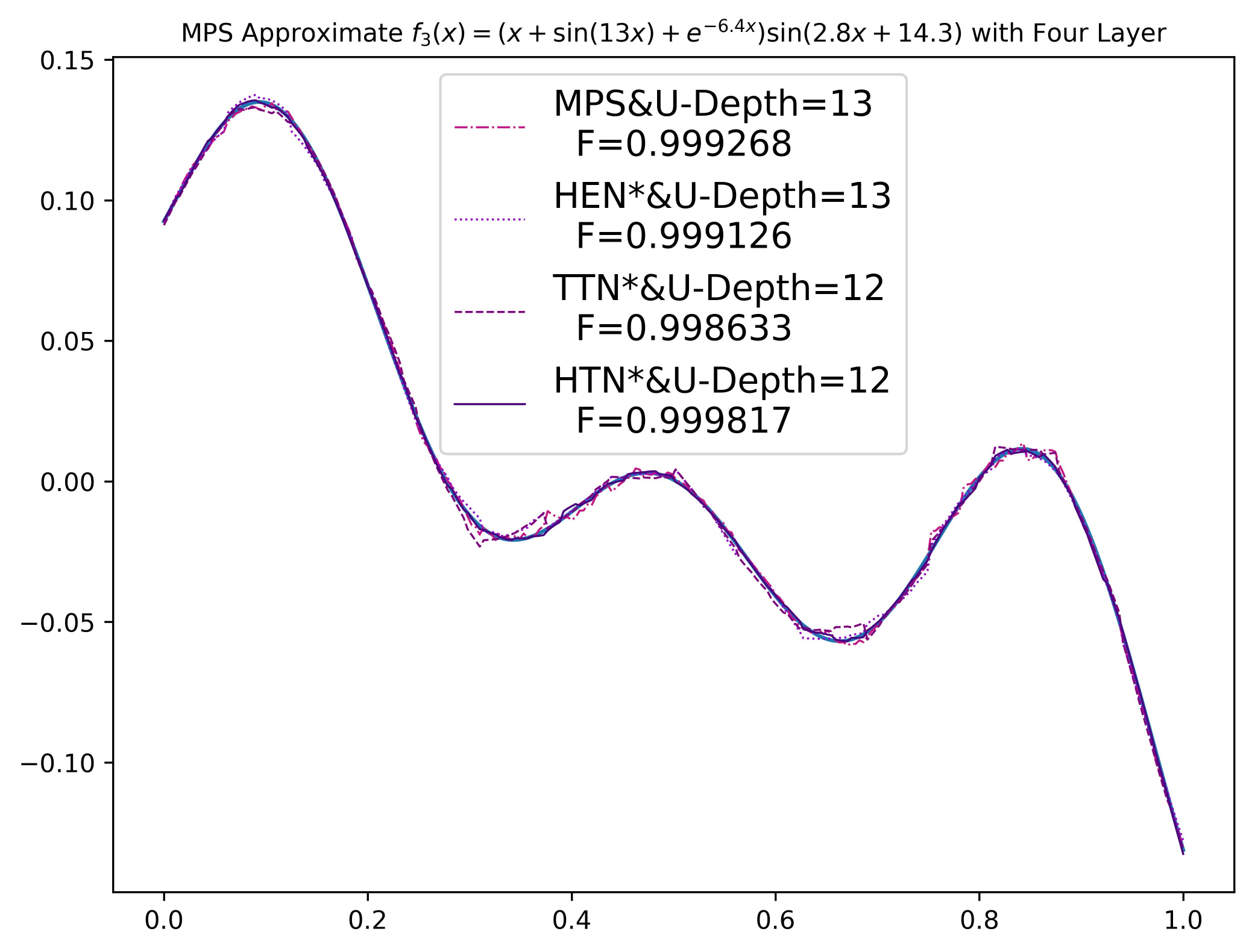}\\[1ex]
		\includegraphics[width=0.23\textwidth]{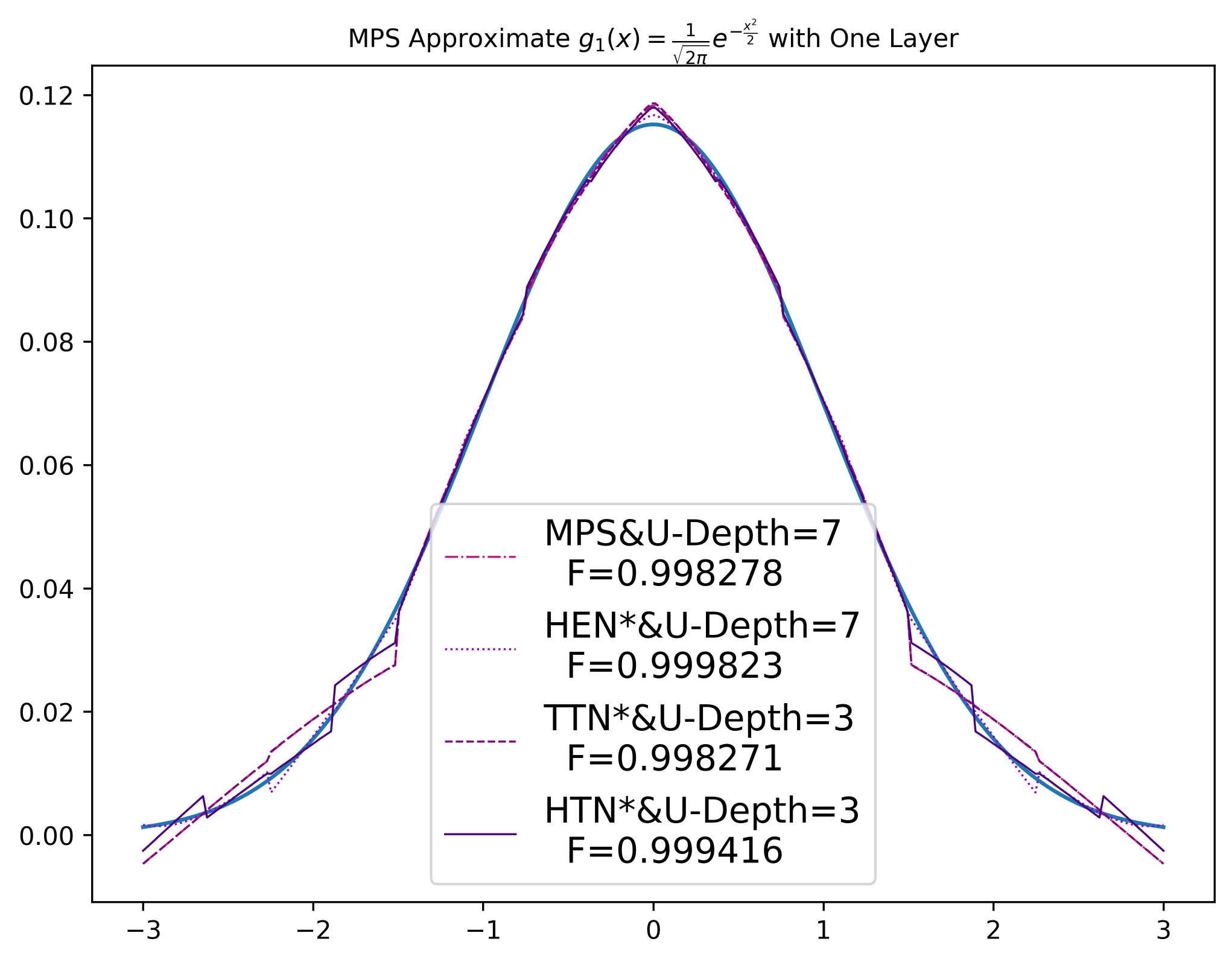}
		\includegraphics[width=0.23\textwidth]{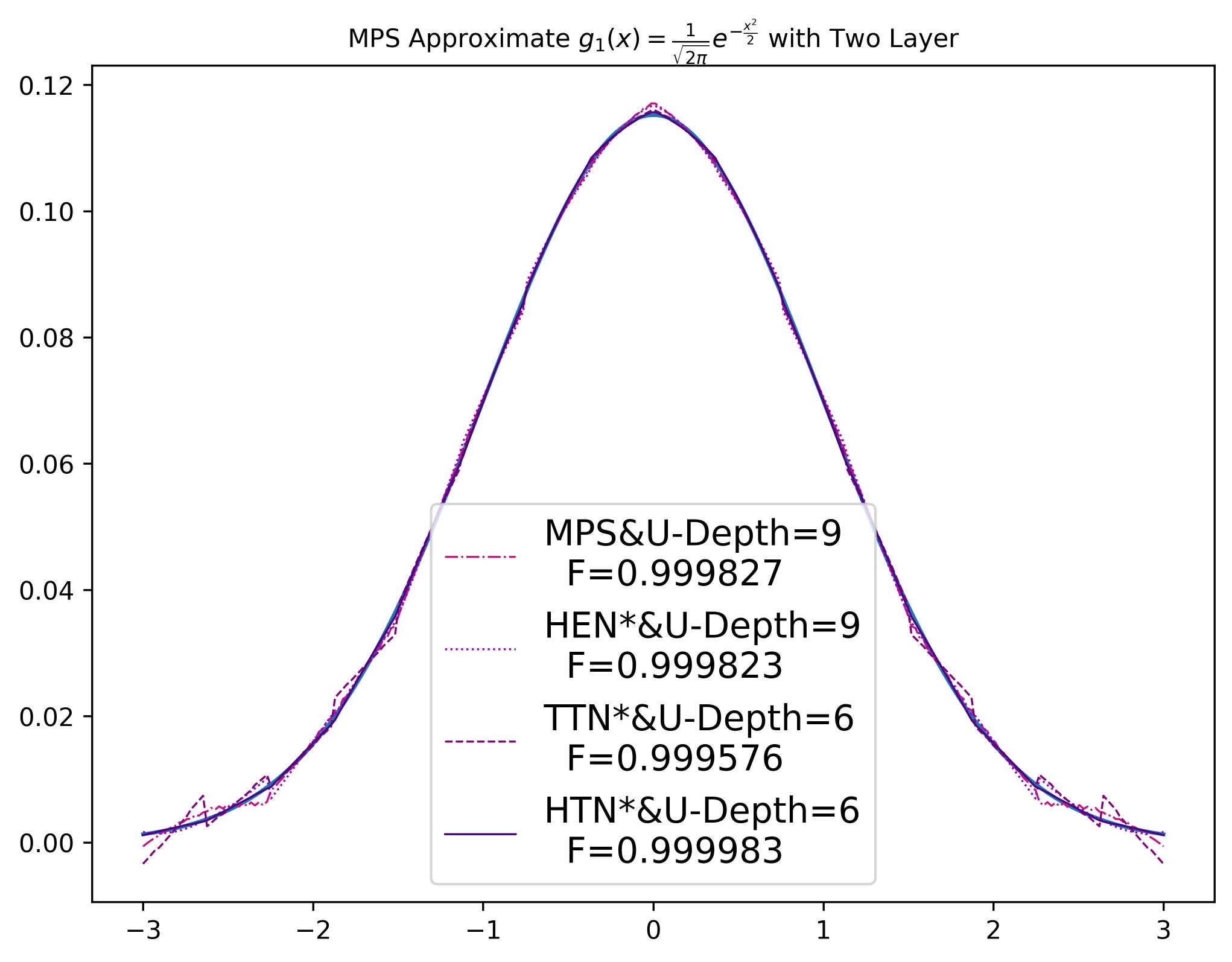}
		\includegraphics[width=0.23\textwidth]{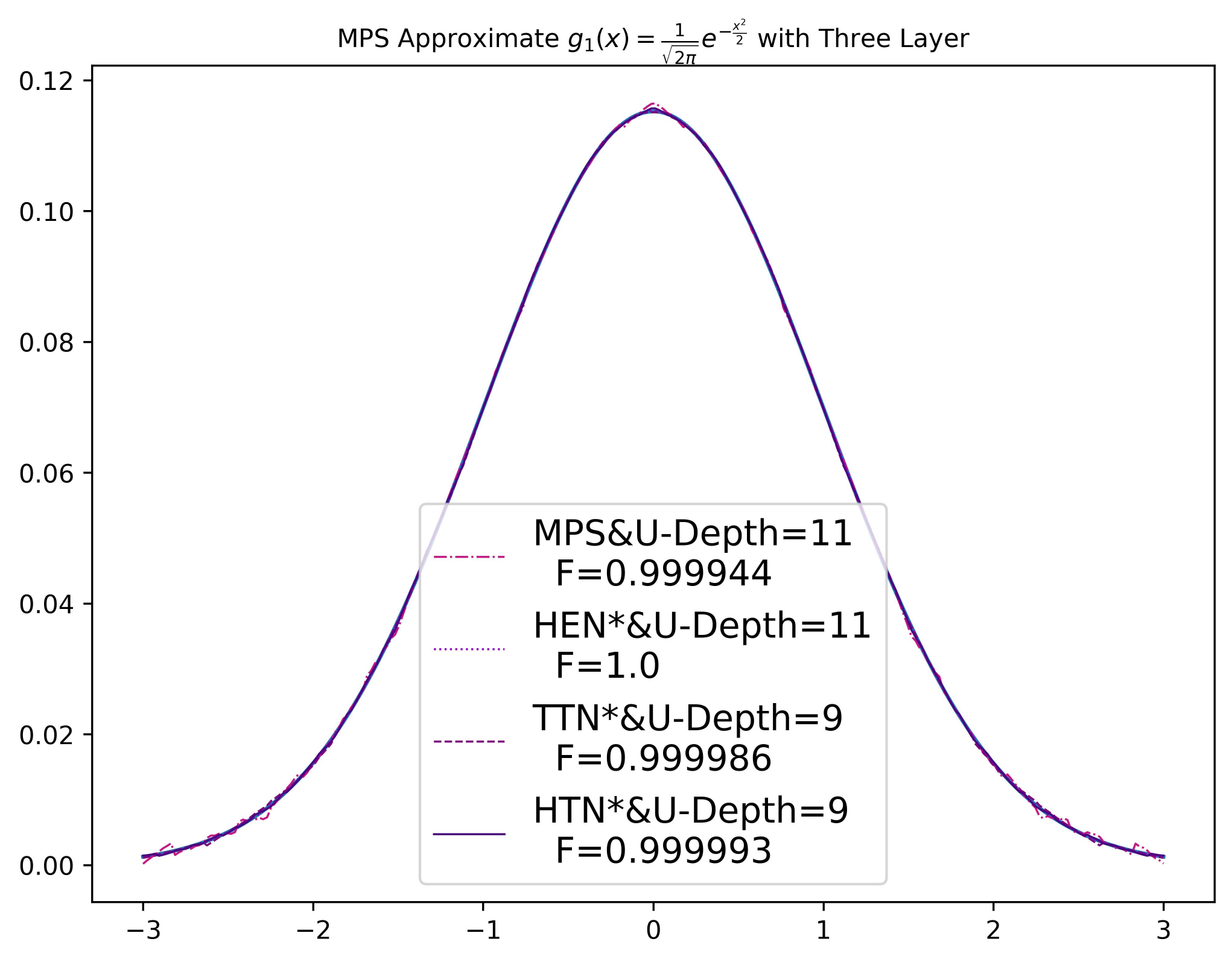}
		\includegraphics[width=0.23\textwidth]{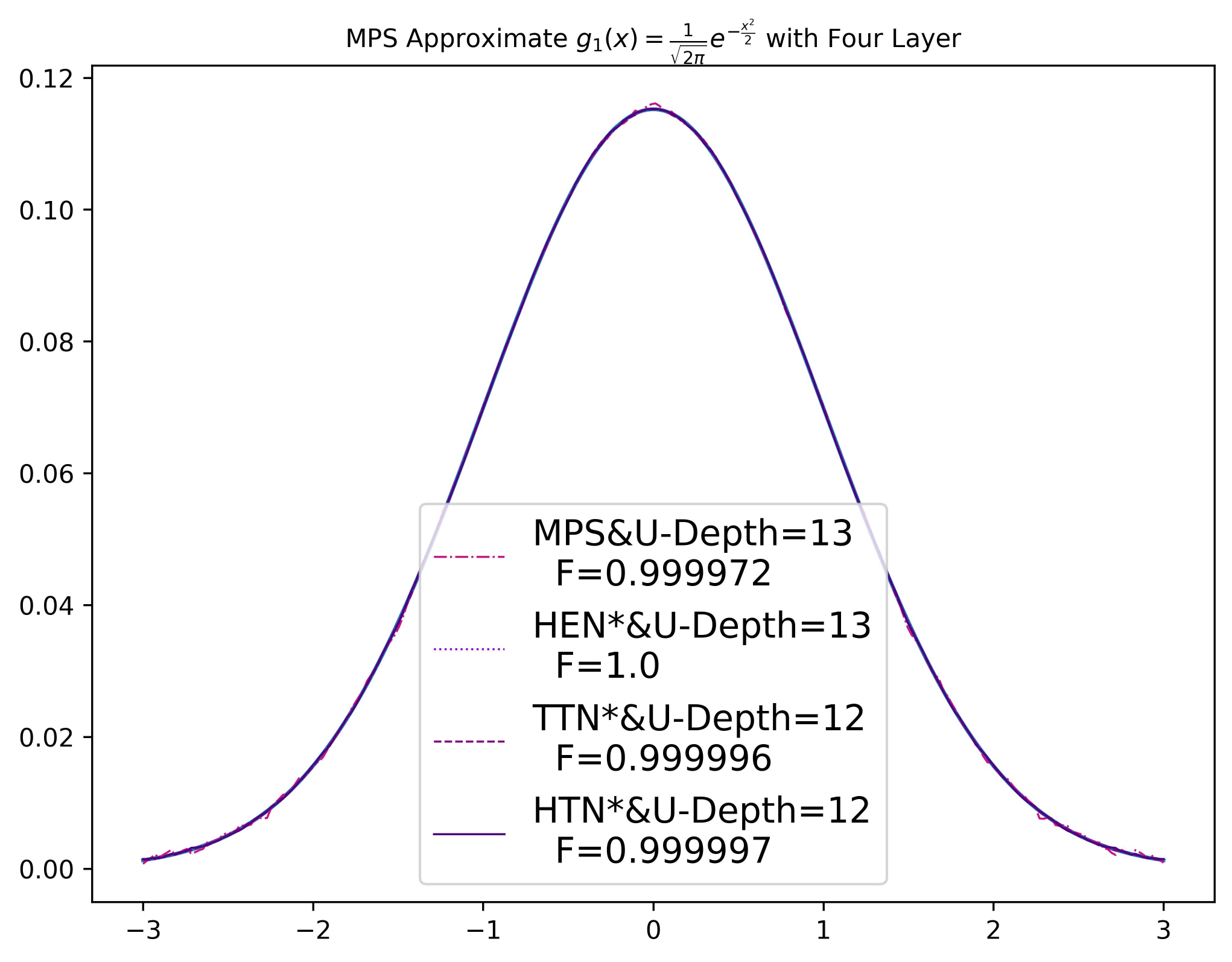}\\[1ex]
		\includegraphics[width=0.23\textwidth]{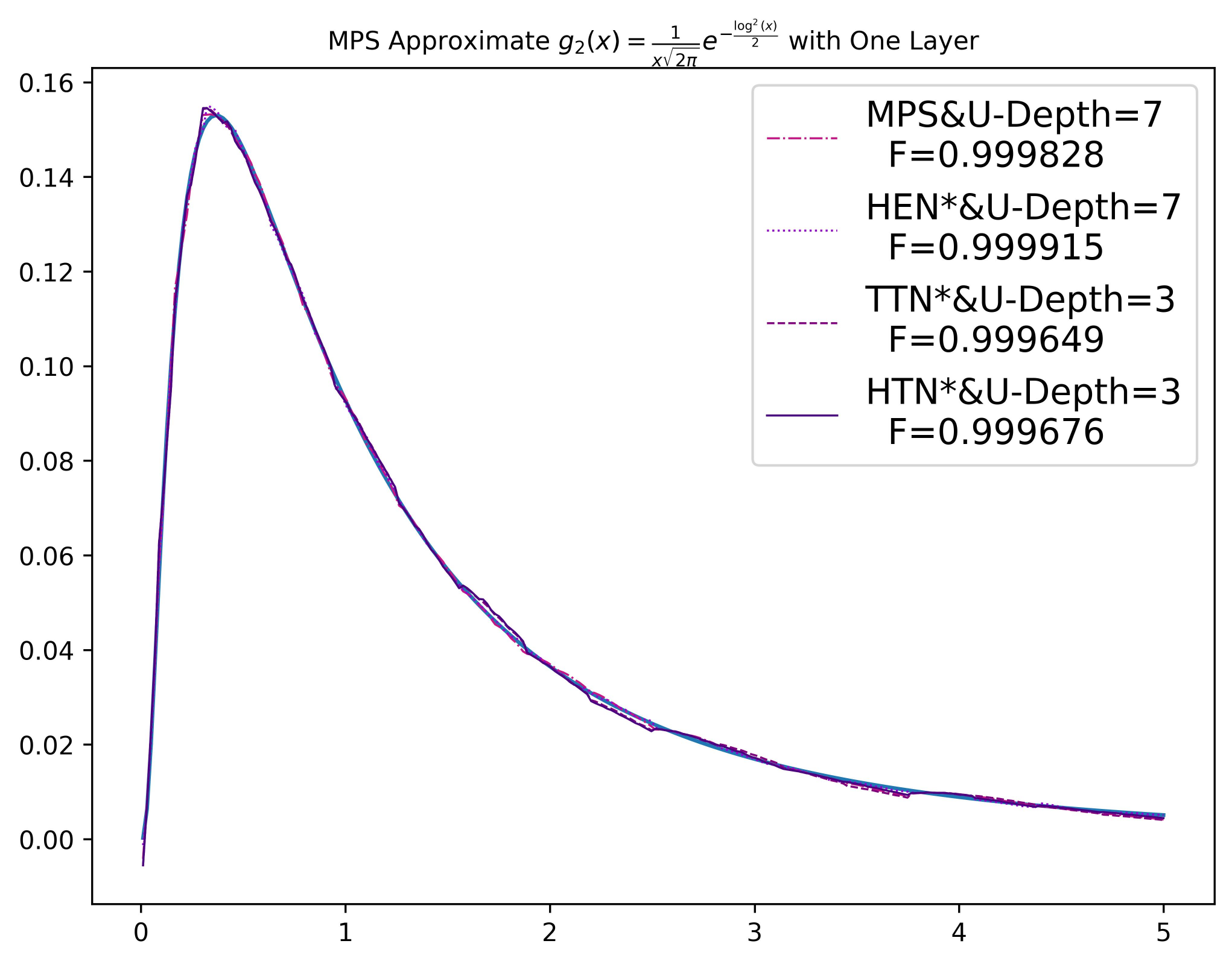}
		\includegraphics[width=0.23\textwidth]{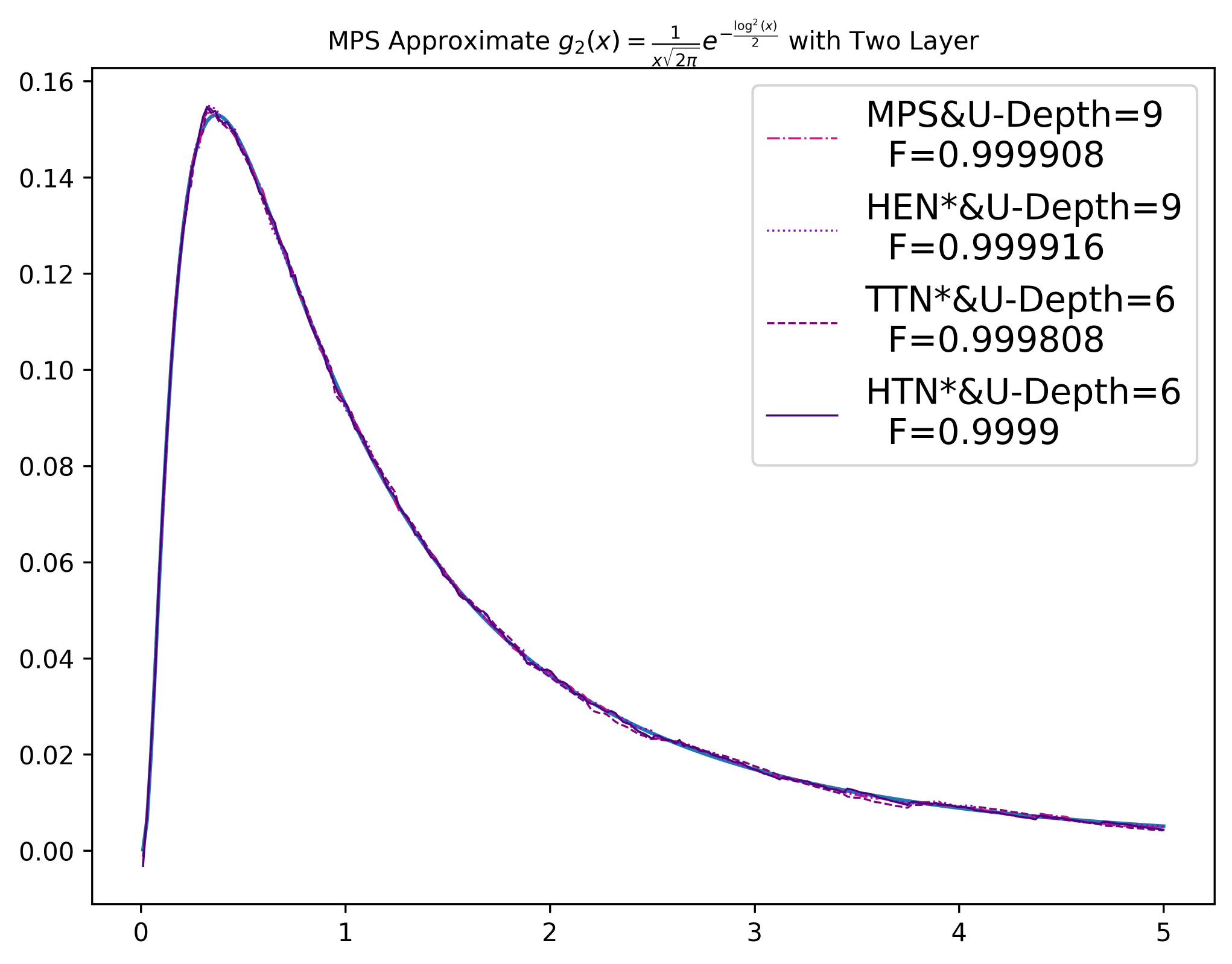}
		\includegraphics[width=0.23\textwidth]{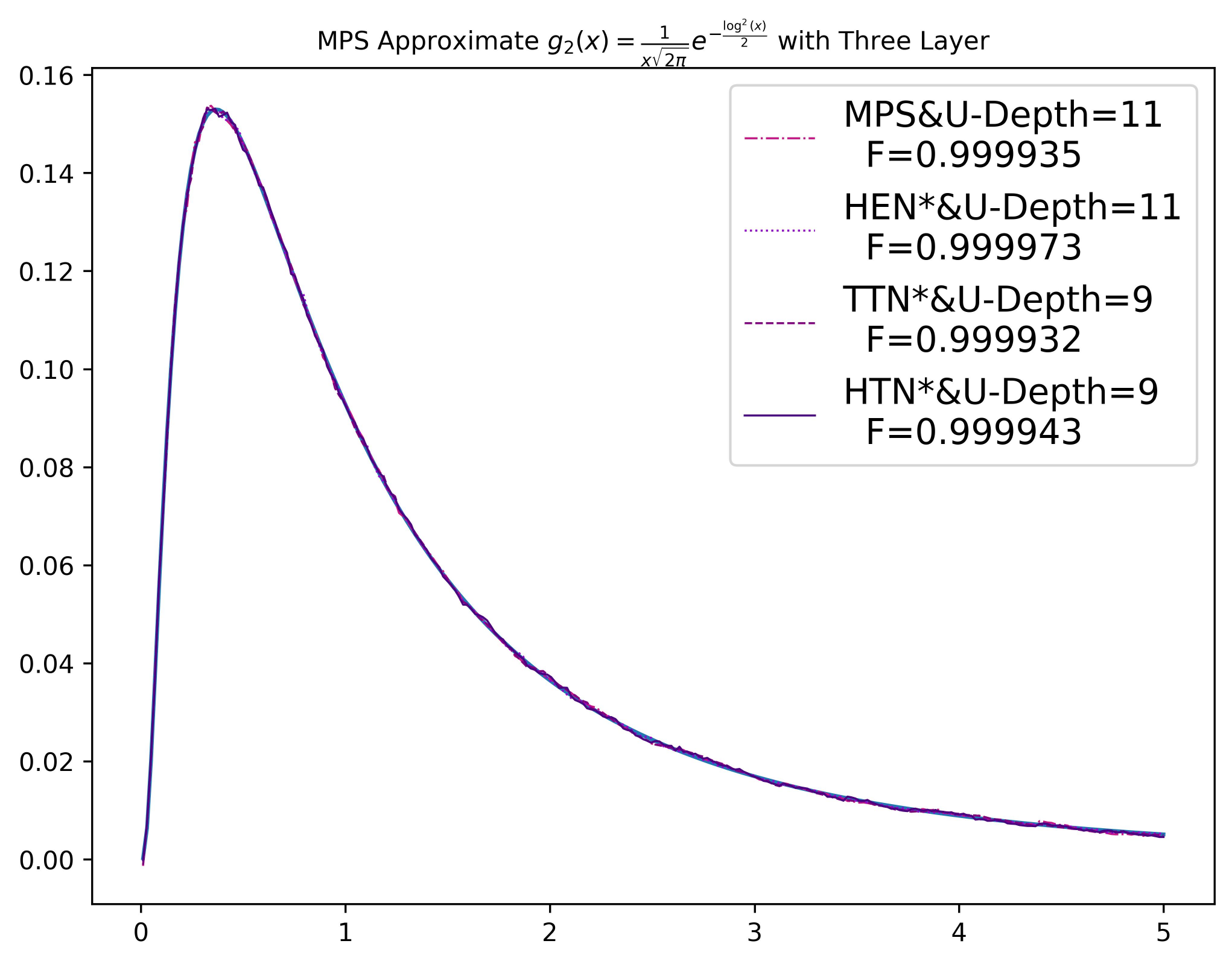}
		\includegraphics[width=0.23\textwidth]{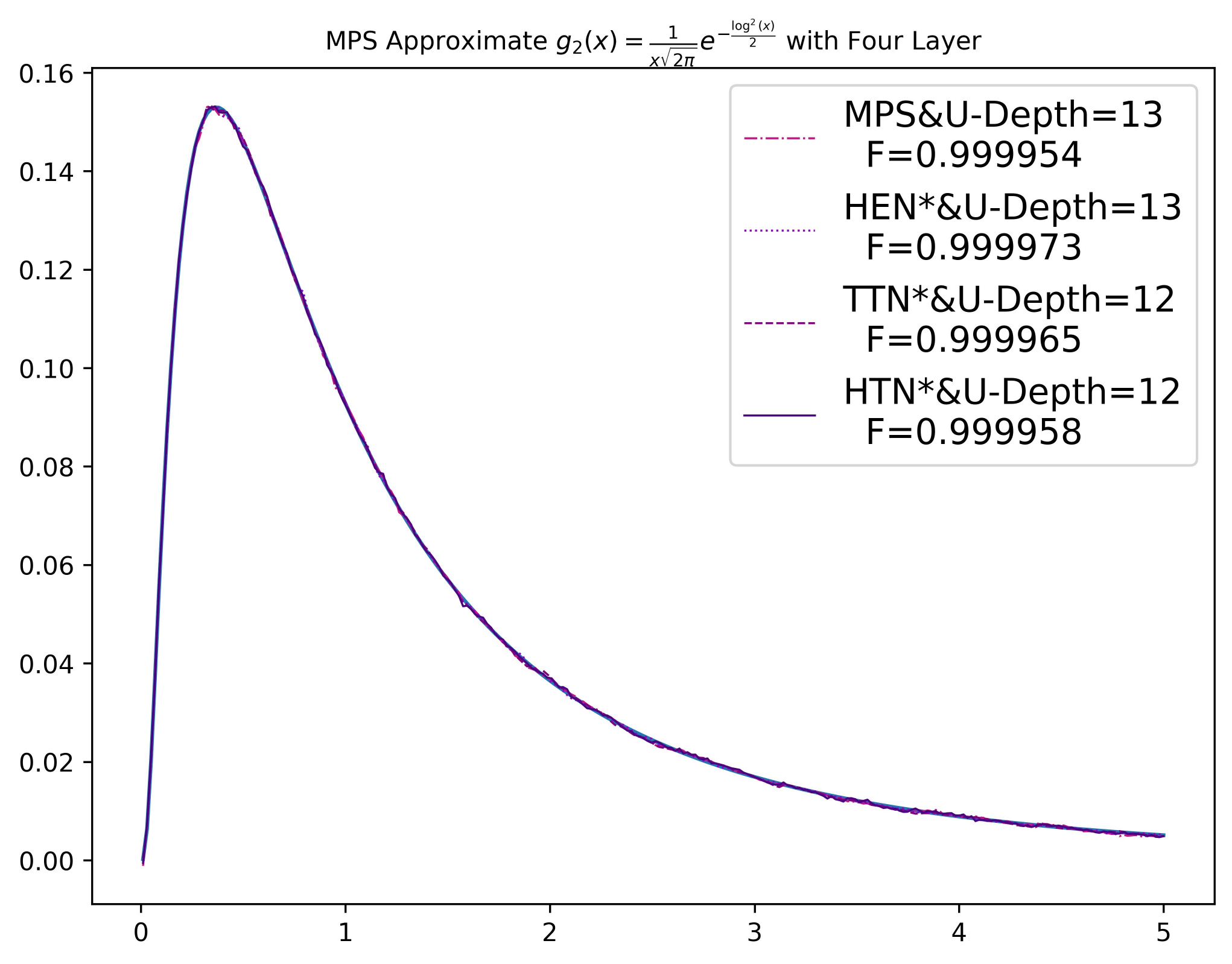}\\[1ex]
		\includegraphics[width=0.23\textwidth]{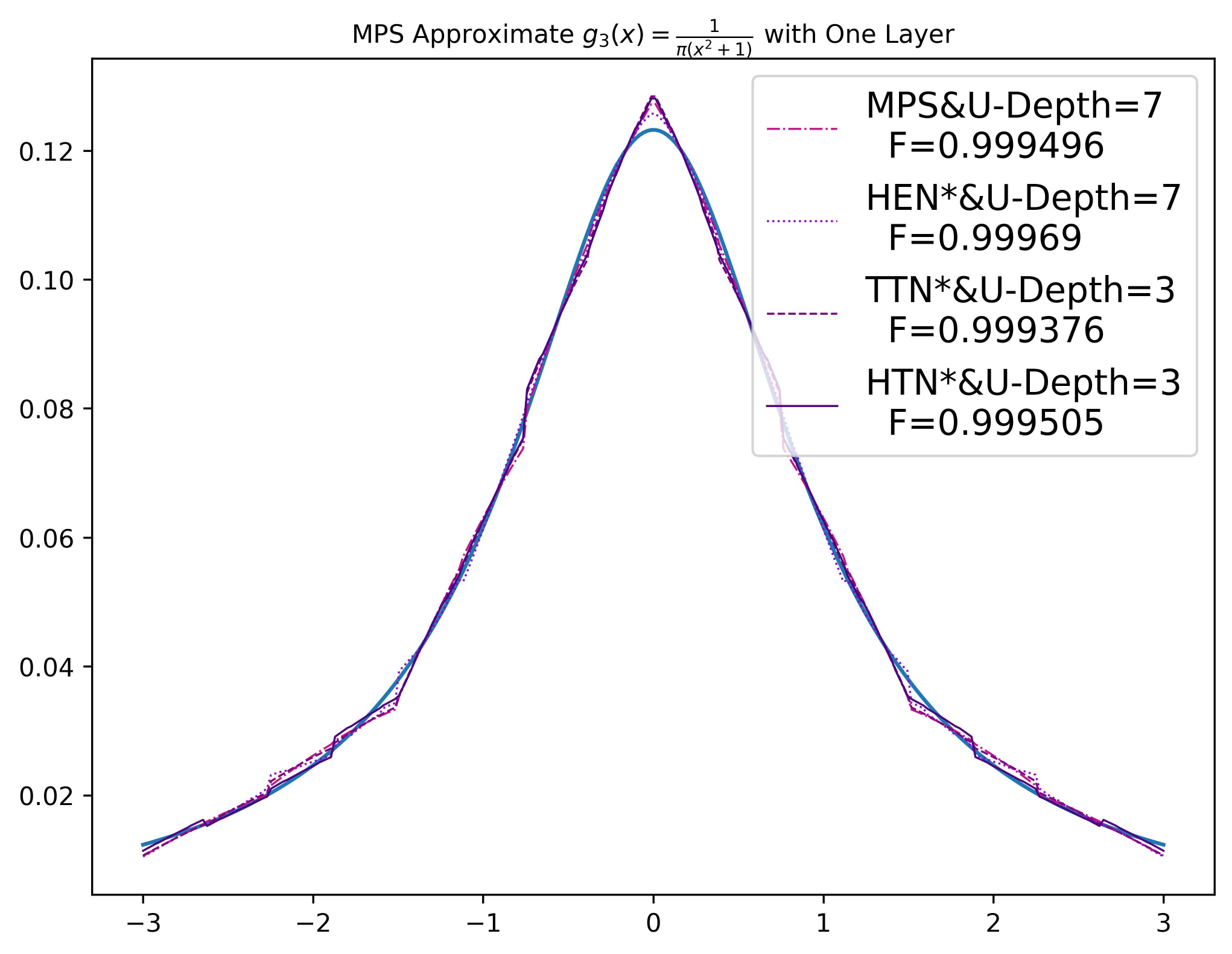}
		\includegraphics[width=0.23\textwidth]{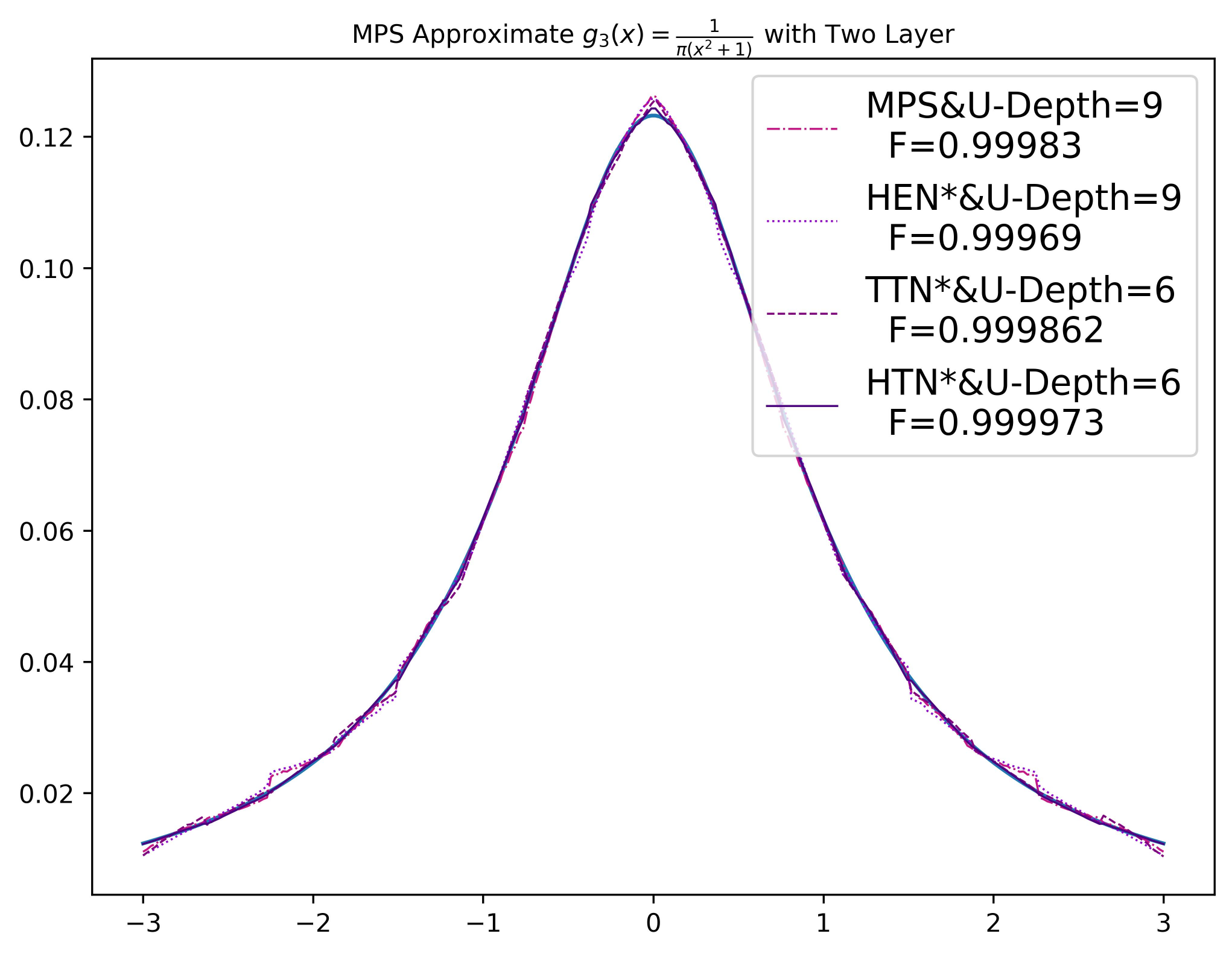}
		\includegraphics[width=0.23\textwidth]{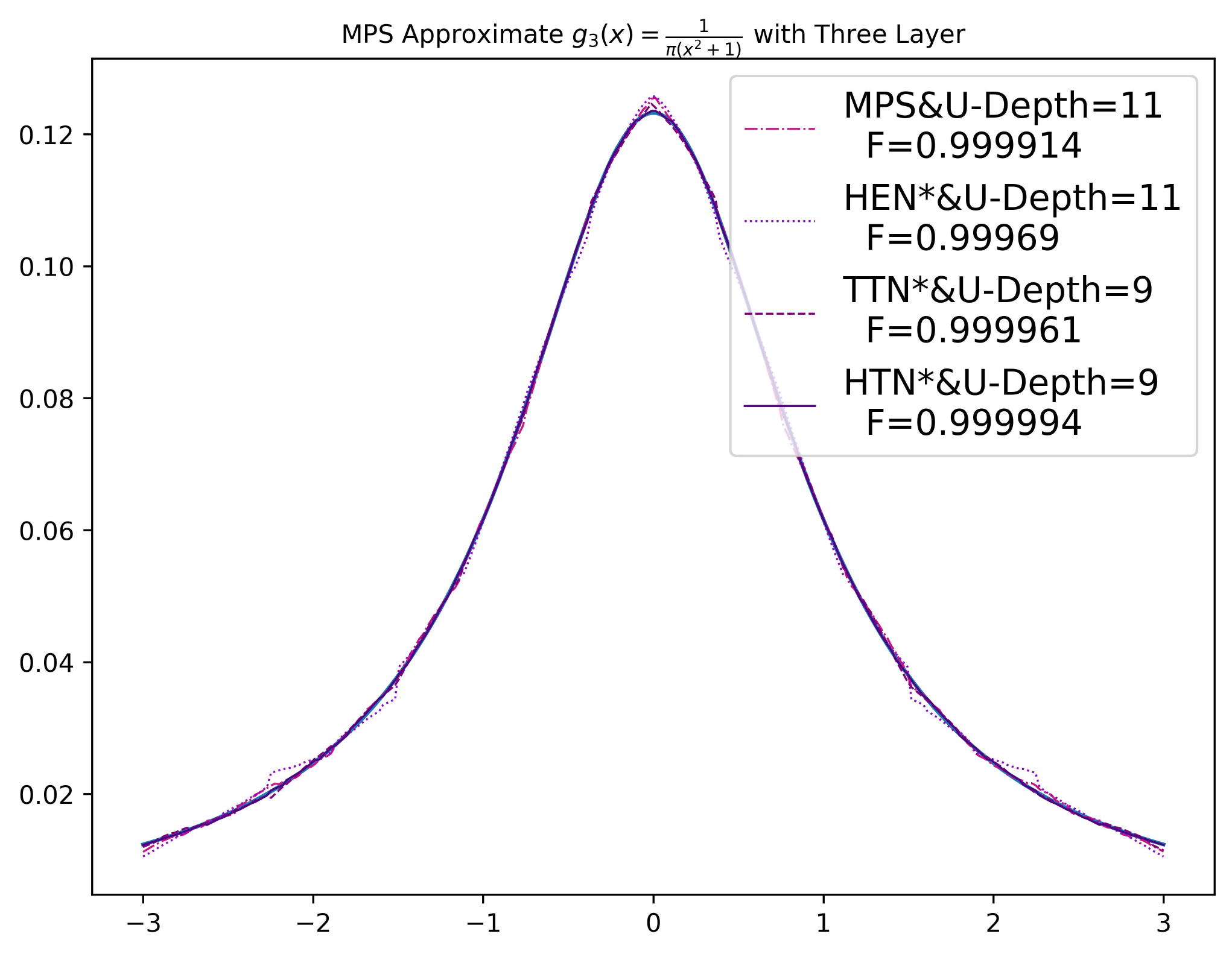}
		\includegraphics[width=0.23\textwidth]{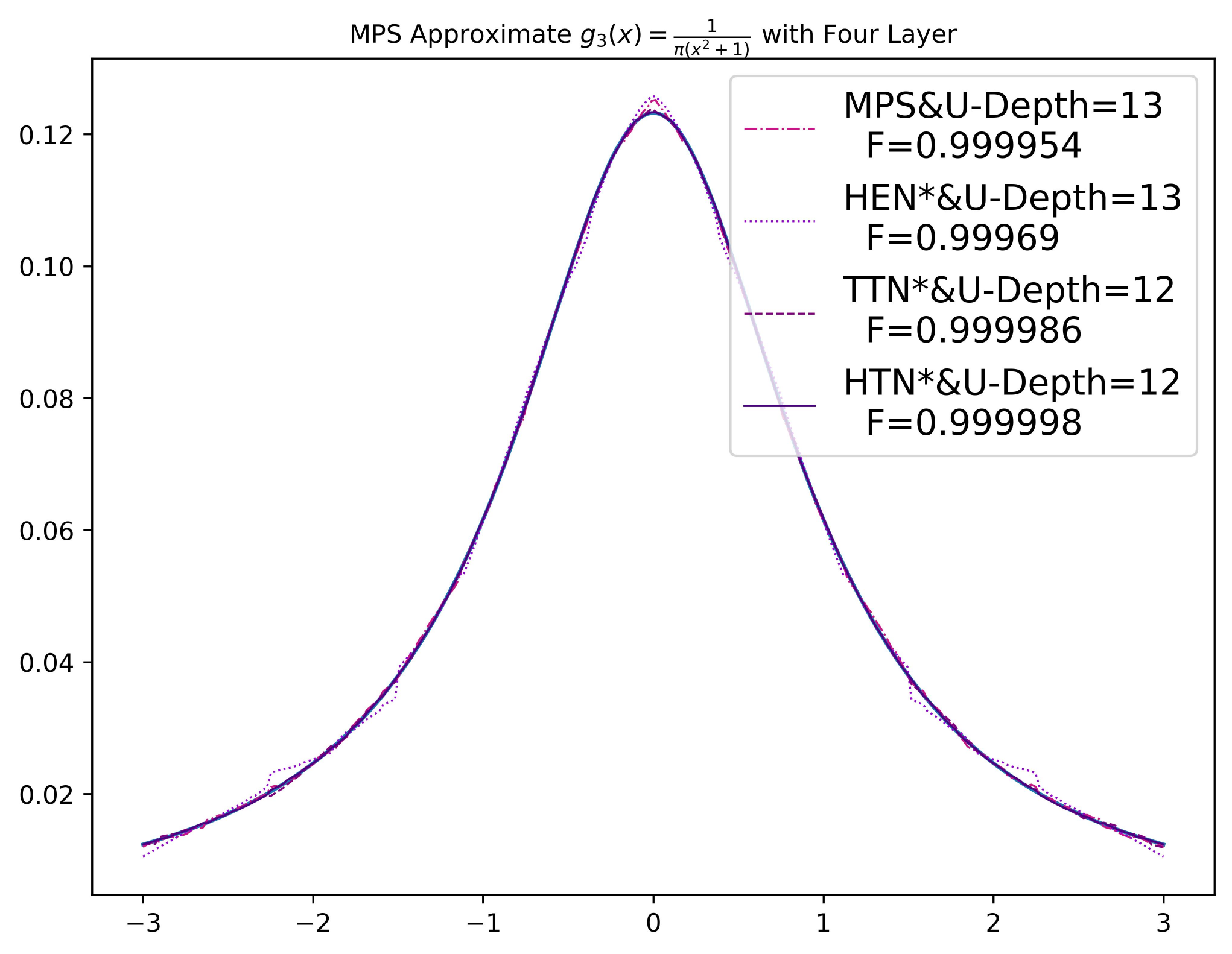}
		\caption{Comparison of amplitude preparation results for the six functions against their original curves, evaluated using the four connectivity schemes at various U-depths.}
		\label{layerfg}
	\end{figure}

To provide a more intuitive illustration of the differences between the original functions and the prepared states, we present a direct comparison between the function curves and the resulting amplitude preparation curves. Furthermore, we investigate the iterative application of the four disentanglement methods to confirm that fidelity improves as the U-depth increases. The specific results are shown in Fig.~\ref{layerfg}.

\begin{figure}[htbp]
		\centering  
		\includegraphics[width=0.75\textwidth]{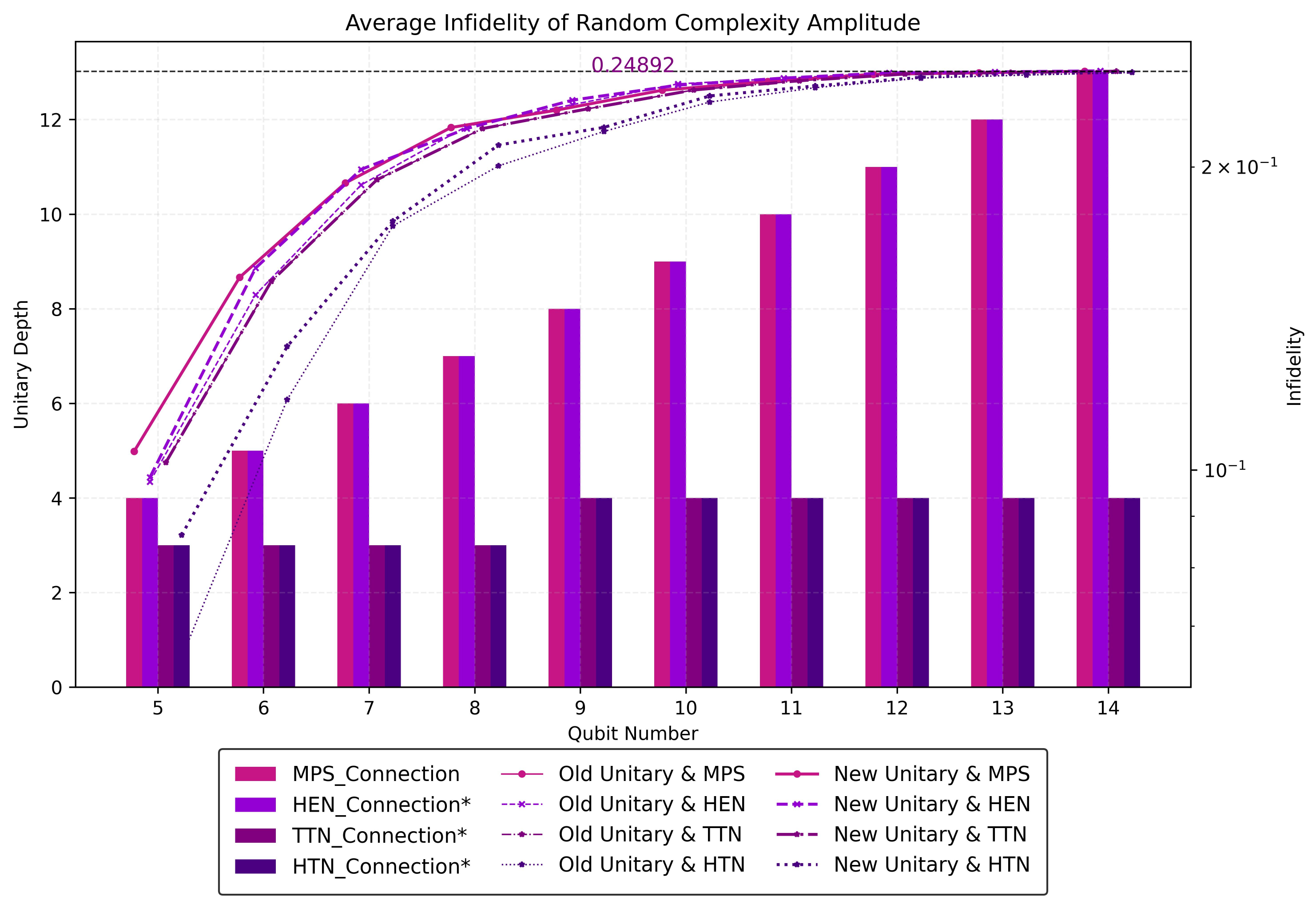}
		\caption{Comparison of the four disentanglement schemes using randomly generated complex amplitudes. Different topological connectivities and the unitary optimization were evaluated. Infidelity values were averaged over 10 random samples per data point to ensure statistical reliability.}
		\label{NewU4}
	\end{figure}

We evaluate the fidelity of the optimized unitary matrix, which requires only two CNOT gates and single-qubit gates, compared to the original matrix that uses three CNOT gates. This evaluation is conducted using the four topological connectivity methods described previously. The complex amplitudes were randomly generated multiple times, and the results were averaged to ensure reliability. As shown in Fig.~\ref{NewU4}, we compare the fidelity as the problem size increases. For each data point, we average the results over 10 random samples. Due to the non-smooth nature of random amplitudes, the fidelity tends to decrease as the scale increases, stabilizing around 12 qubits. All our experiments in this test used only one layer, although hardware implementations might involve multiple layers.

For a chip with a specific topology, we consider an $n \times m$ grid, which is common in superconducting quantum computers. With our proposed scheme, the quantum state can be contracted simultaneously from the two outer rows or columns. Each contraction step disentangles approximately $2 \cdot \max(m, n)$ qubits, requiring a U-depth of about $\min(m, n)/2$. Once the state's weight is contracted to a single line, simultaneous contractions from both ends of the line can disentangle 2 qubits per step, necessitating a U-depth of approximately $\max(m, n)/2$. Therefore, amplitude preparation for $nm$ qubits on an $n \times m$ grid chip requires a circuit depth of approximately $(m+n)/2$. When $m$ and $n$ are comparable, this corresponds to a circuit depth on the order of the square root of the number of qubits. We can also determine that functions with an MPS rank of 2 or less, as mentioned previously, can be prepared exactly on an $n \times m$ grid chip with a circuit depth of $O(\max(m, n))$.
	
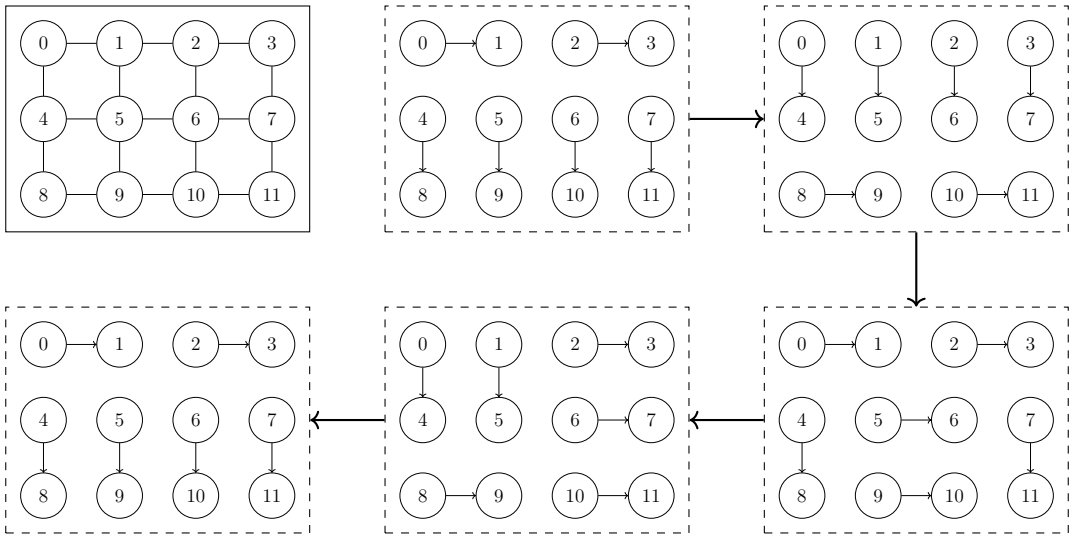
\begin{figure}[H]
	\centering
	\resizebox{0.95\linewidth}{!}{
	\begin{tikzpicture}
		% 第一排 (三个子图)
		\node[draw, rectangle, minimum width=3cm, minimum height=3cm] (A) at (1, 0) {
			\resizebox{0.25\linewidth}{!}{
				% 第一个子图
				\begin{tikzpicture}[scale=2, minimum size=1.2cm, font=\Large]
					\node[draw, circle] (A) at (0,2) {0};
					\node[draw, circle] (B) at (1,2) {1};
					\node[draw, circle] (C) at (2,2) {2};
					\node[draw, circle] (D) at (3,2) {3};
					\node[draw, circle] (E) at (0,1) {4};
					\node[draw, circle] (F) at (1,1) {5};
					\node[draw, circle] (G) at (2,1) {6};
					\node[draw, circle] (H) at (3,1) {7};
					\node[draw, circle] (I) at (0,0) {8};
					\node[draw, circle] (J) at (1,0) {9};
					\node[draw, circle] (K) at (2,0) {10};
					\node[draw, circle] (L) at (3,0) {11};
					\draw[-] (A) -- (B);
					\draw[-] (B) -- (C);
					\draw[-] (C) -- (D);
					\draw[-] (E) -- (F);
					\draw[-] (F) -- (G);
					\draw[-] (G) -- (H);
					\draw[-] (I) -- (J);
					\draw[-] (J) -- (K);
					\draw[-] (K) -- (L);
					\draw[-] (A) -- (E);
					\draw[-] (E) -- (I);
					\draw[-] (B) -- (F);
					\draw[-] (F) -- (J);
					\draw[-] (C) -- (G);
					\draw[-] (G) -- (K);
					\draw[-] (D) -- (H);
					\draw[-] (H) -- (L);
				\end{tikzpicture}
			}
		};

		\node[draw, dashed, rectangle, minimum width=3cm, minimum height=3cm] (A) at (6, 0) {
			\resizebox{0.25\linewidth}{!}{
				% 第一个子图
				\begin{tikzpicture}[scale=2, minimum size=1.2cm, font=\Large]
					\node[draw, circle, solid] (0) at (0,2) {0};
					\node[draw, circle, solid] (1) at (1,2) {1};
					\node[draw, circle, solid] (2) at (2,2) {2};
					\node[draw, circle, solid] (3) at (3,2) {3};
					\node[draw, circle, solid] (4) at (0,1) {4};
					\node[draw, circle, solid] (5) at (1,1) {5};
					\node[draw, circle, solid] (6) at (2,1) {6};
					\node[draw, circle, solid] (7) at (3,1) {7};
					\node[draw, circle, solid] (8) at (0,0) {8};
					\node[draw, circle, solid] (9) at (1,0) {9};
					\node[draw, circle, solid] (10) at (2,0) {10};
					\node[draw, circle, solid] (11) at (3,0) {11};
					\draw[->, solid] (0) -- (1);
					\draw[->, solid] (2) -- (3);
					\draw[->, solid] (4) -- (8);
					\draw[->, solid] (5) -- (9);
					\draw[->, solid] (6) -- (10);
					\draw[->, solid] (7) -- (11);
				\end{tikzpicture}
			}
		};
		
		\node[draw, dashed, rectangle, minimum width=3cm, minimum height=3cm] (B) at (11, 0) {
			\resizebox{0.25\linewidth}{!}{
				% 第二个子图
				\begin{tikzpicture}[scale=2, minimum size=1.2cm, font=\Large]
					\node[draw, circle, solid] (0) at (0,2) {0};
					\node[draw, circle, solid] (1) at (1,2) {1};
					\node[draw, circle, solid] (2) at (2,2) {2};
					\node[draw, circle, solid] (3) at (3,2) {3};
					\node[draw, circle, solid] (4) at (0,1) {4};
					\node[draw, circle, solid] (5) at (1,1) {5};
					\node[draw, circle, solid] (6) at (2,1) {6};
					\node[draw, circle, solid] (7) at (3,1) {7};
					\node[draw, circle, solid] (8) at (0,0) {8};
					\node[draw, circle, solid] (9) at (1,0) {9};
					\node[draw, circle, solid] (10) at (2,0) {10};
					\node[draw, circle, solid] (11) at (3,0) {11};
					\draw[->, solid] (0) -- (4);
					\draw[->, solid] (1) -- (5);
					\draw[->, solid] (2) -- (6);
					\draw[->, solid] (3) -- (7);
					\draw[->, solid] (8) -- (9);
					\draw[->, solid] (10) -- (11);
				\end{tikzpicture}
			}
		};
		
		\node[draw, dashed, rectangle, minimum width=3cm, minimum height=3cm] (C) at (11, -4) {
			\resizebox{0.25\linewidth}{!}{
				% 第三个子图
				\begin{tikzpicture}[scale=2, minimum size=1.2cm, font=\Large]
					\node[draw, circle, solid] (0) at (0,2) {0};
					\node[draw, circle, solid] (1) at (1,2) {1};
					\node[draw, circle, solid] (2) at (2,2) {2};
					\node[draw, circle, solid] (3) at (3,2) {3};
					\node[draw, circle, solid] (4) at (0,1) {4};
					\node[draw, circle, solid] (5) at (1,1) {5};
					\node[draw, circle, solid] (6) at (2,1) {6};
					\node[draw, circle, solid] (7) at (3,1) {7};
					\node[draw, circle, solid] (8) at (0,0) {8};
					\node[draw, circle, solid] (9) at (1,0) {9};
					\node[draw, circle, solid] (10) at (2,0) {10};
					\node[draw, circle, solid] (11) at (3,0) {11};
					\draw[->, solid] (0) -- (1);
					\draw[->, solid] (2) -- (3);
					\draw[->, solid] (4) -- (8);
					\draw[->, solid] (5) -- (6);
					\draw[->, solid] (9) -- (10);
					\draw[->, solid] (7) -- (11);
				\end{tikzpicture}
			}
		};
		
		% 第一排箭头
		\draw[->, thick] (A) -- (B);
		\draw[->, thick] (B) -- (C);
		
		% 第二排 (第四个和第五个子图)
		\node[draw, dashed, rectangle, minimum width=3cm, minimum height=3cm] (D) at (1, -4) {
			\resizebox{0.25\linewidth}{!}{
				% 第四个子图
				\begin{tikzpicture}[scale=2, minimum size=1.2cm, font=\Large]
					\node[draw, circle, solid] (0) at (0,2) {0};
					\node[draw, circle, solid] (1) at (1,2) {1};
					\node[draw, circle, solid] (2) at (2,2) {2};
					\node[draw, circle, solid] (3) at (3,2) {3};
					\node[draw, circle, solid] (4) at (0,1) {4};
					\node[draw, circle, solid] (5) at (1,1) {5};
					\node[draw, circle, solid] (6) at (2,1) {6};
					\node[draw, circle, solid] (7) at (3,1) {7};
					\node[draw, circle, solid] (8) at (0,0) {8};
					\node[draw, circle, solid] (9) at (1,0) {9};
					\node[draw, circle, solid] (10) at (2,0) {10};
					\node[draw, circle, solid] (11) at (3,0) {11};
					\draw[->, solid] (0) -- (1);
					\draw[->, solid] (2) -- (3);
					\draw[->, solid] (4) -- (8);
					\draw[->, solid] (5) -- (9);
					\draw[->, solid] (6) -- (10);
					\draw[->, solid] (7) -- (11);
				\end{tikzpicture}
			}
		};
		
		\node[draw, dashed, rectangle, minimum width=3cm, minimum height=3cm] (E) at (6, -4) {
			\resizebox{0.25\linewidth}{!}{
				% 第五个子图
				\begin{tikzpicture}[scale=2, minimum size=1.2cm, font=\Large]
					\node[draw, circle, solid] (0) at (0,2) {0};
					\node[draw, circle, solid] (1) at (1,2) {1};
					\node[draw, circle, solid] (2) at (2,2) {2};
					\node[draw, circle, solid] (3) at (3,2) {3};
					\node[draw, circle, solid] (4) at (0,1) {4};
					\node[draw, circle, solid] (5) at (1,1) {5};
					\node[draw, circle, solid] (6) at (2,1) {6};
					\node[draw, circle, solid] (7) at (3,1) {7};
					\node[draw, circle, solid] (8) at (0,0) {8};
					\node[draw, circle, solid] (9) at (1,0) {9};
					\node[draw, circle, solid] (10) at (2,0) {10};
					\node[draw, circle, solid] (11) at (3,0) {11};
					\draw[->, solid] (0) -- (4);
					\draw[->, solid] (1) -- (5);
					\draw[->, solid] (8) -- (9);
					\draw[->, solid] (2) -- (3);
					\draw[->, solid] (6) -- (7);
					\draw[->, solid] (10) -- (11);
				\end{tikzpicture}
			}
		};
		% 添加向下箭头连接C和E
		\draw[->, thick] (C) -- (E);
		\draw[<-, thick] (D) -- (E);
	\end{tikzpicture}
	}
	\caption{Proposed connectivity scheme for MPS amplitude preparation on a 3-row, 4-column grid-based chip topology, where parallel fourth-order unitary operations effectively reduce circuit depth.}
	\label{topway}
\end{figure}

To validate our method's applicability to common functions on a grid chip topology, we provide a concrete experimental example using a $3 \times 4$ grid structure. As illustrated in Fig.~\ref{topway}, we do not follow the aforementioned bidirectional contraction scheme; instead, we showcase a custom scheme to demonstrate the flexibility of our approach. We use arrows to represent the disentanglement method, which can be intuitively understood as transferring the quantum state's weight to the next qubit along the arrow's direction. In our scheme, we reset the amplitude at each U-depth. Compared to the traditional MPS scheme, which requires a U-depth of 11, our method achieves higher fidelity at U-depths of 5 and 10. Furthermore, we tested the fidelity of the HEN scheme, and the results confirm the comprehensive superiority of our method.

\begin{figure}[htbp]
		\centering  
		\includegraphics[width=0.84\textwidth]{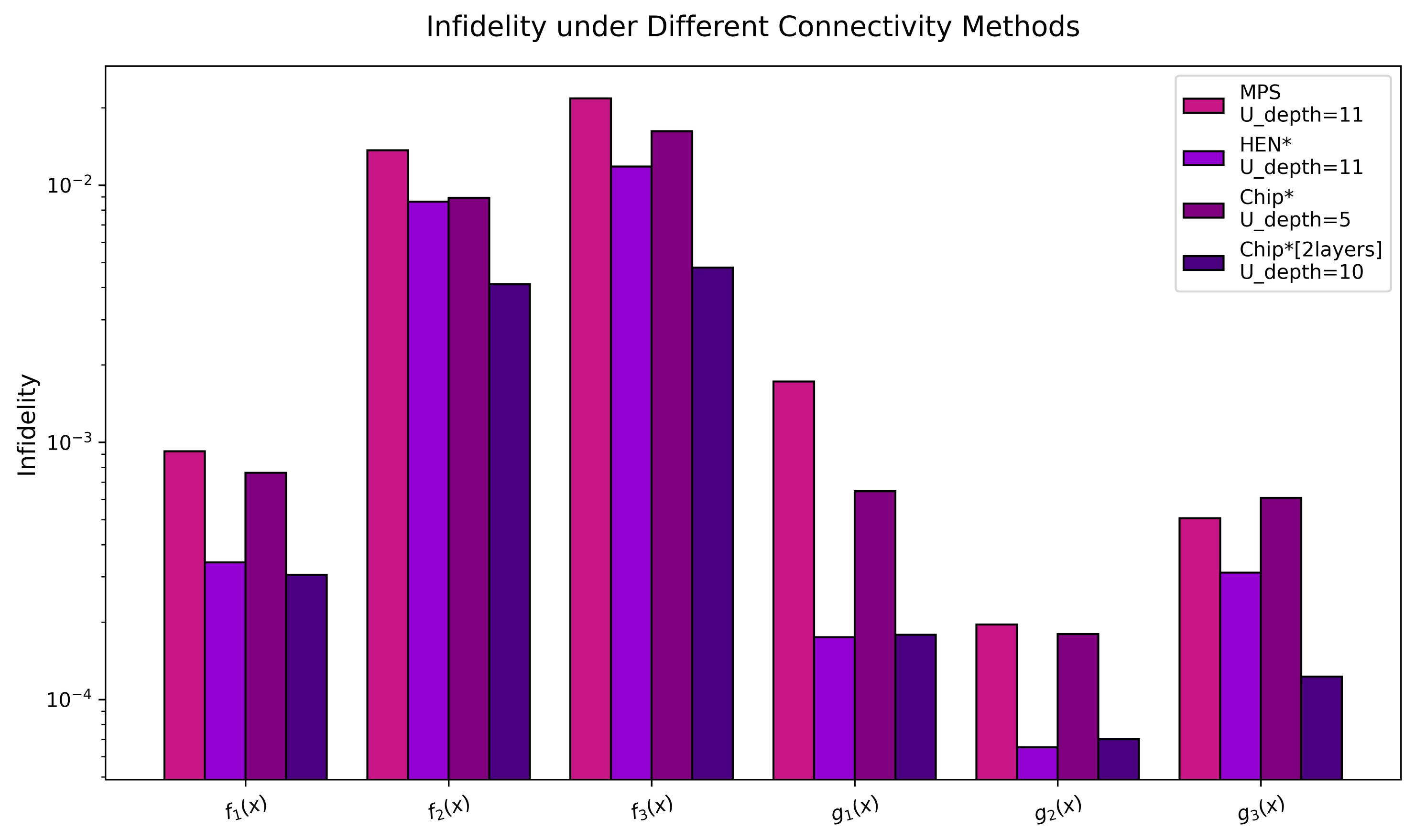}
		\caption{Comparison of infidelity for the six functions between the traditional MPS and the chip-adapted IMPS. The chip-adapted method reduces the circuit depth for 12 qubits to U-depths of 5 and 10, whereas the traditional method has a U-depth of 11.}
		\label{Chiptop}
	\end{figure}

\begin{figure}[htbp]
		\centering
		\includegraphics[width=0.3\textwidth]{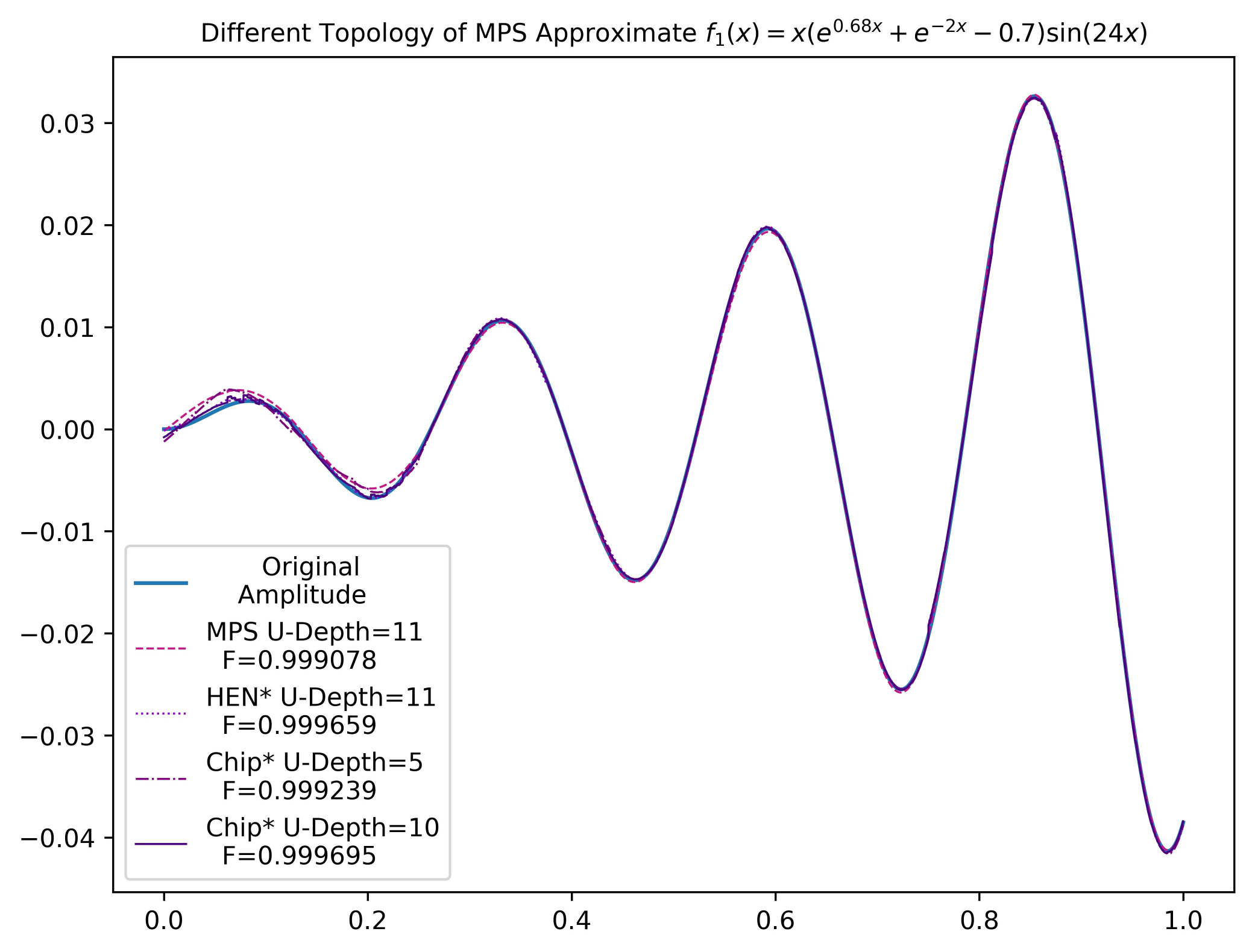}
		\includegraphics[width=0.3\textwidth]{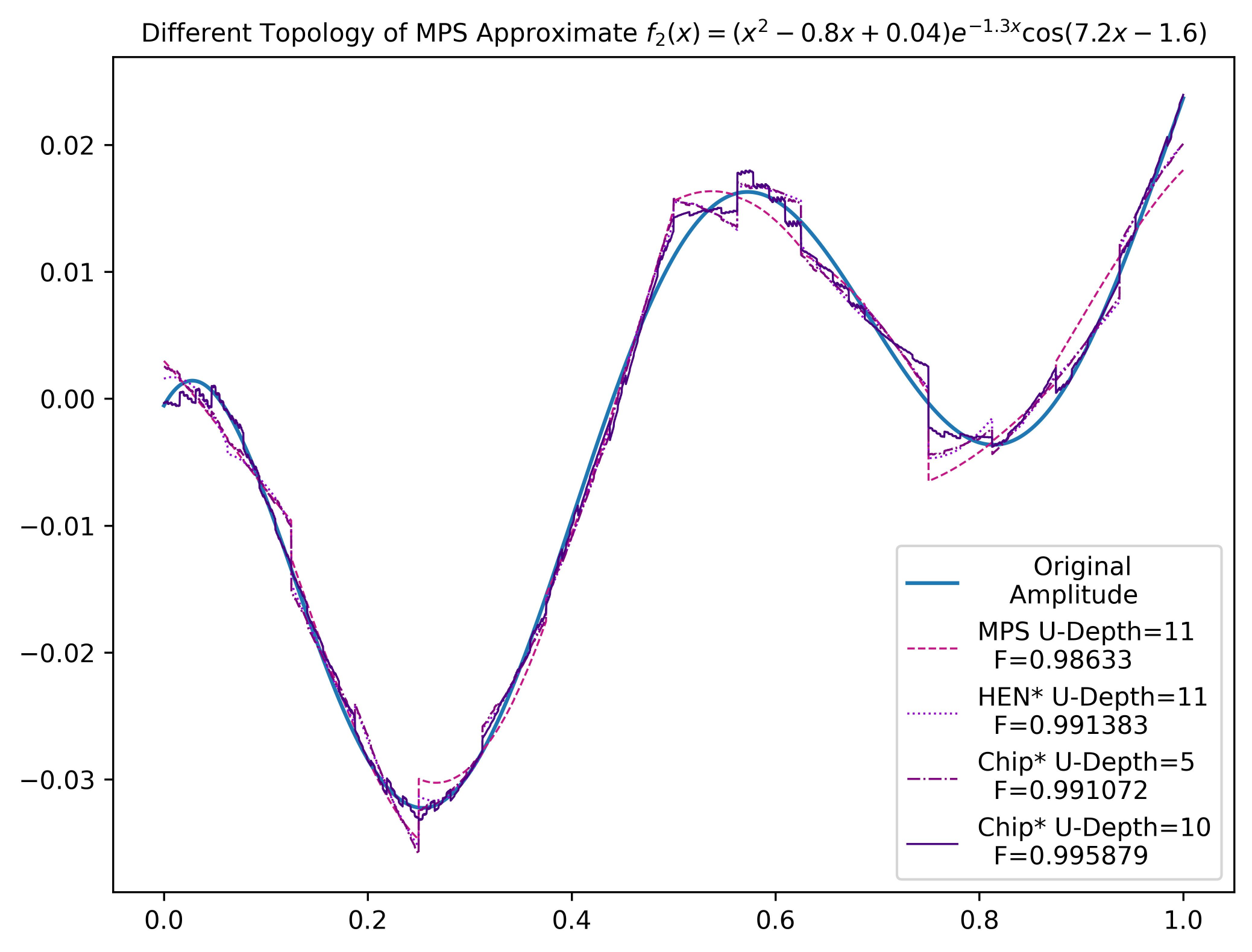}
		\includegraphics[width=0.3\textwidth]{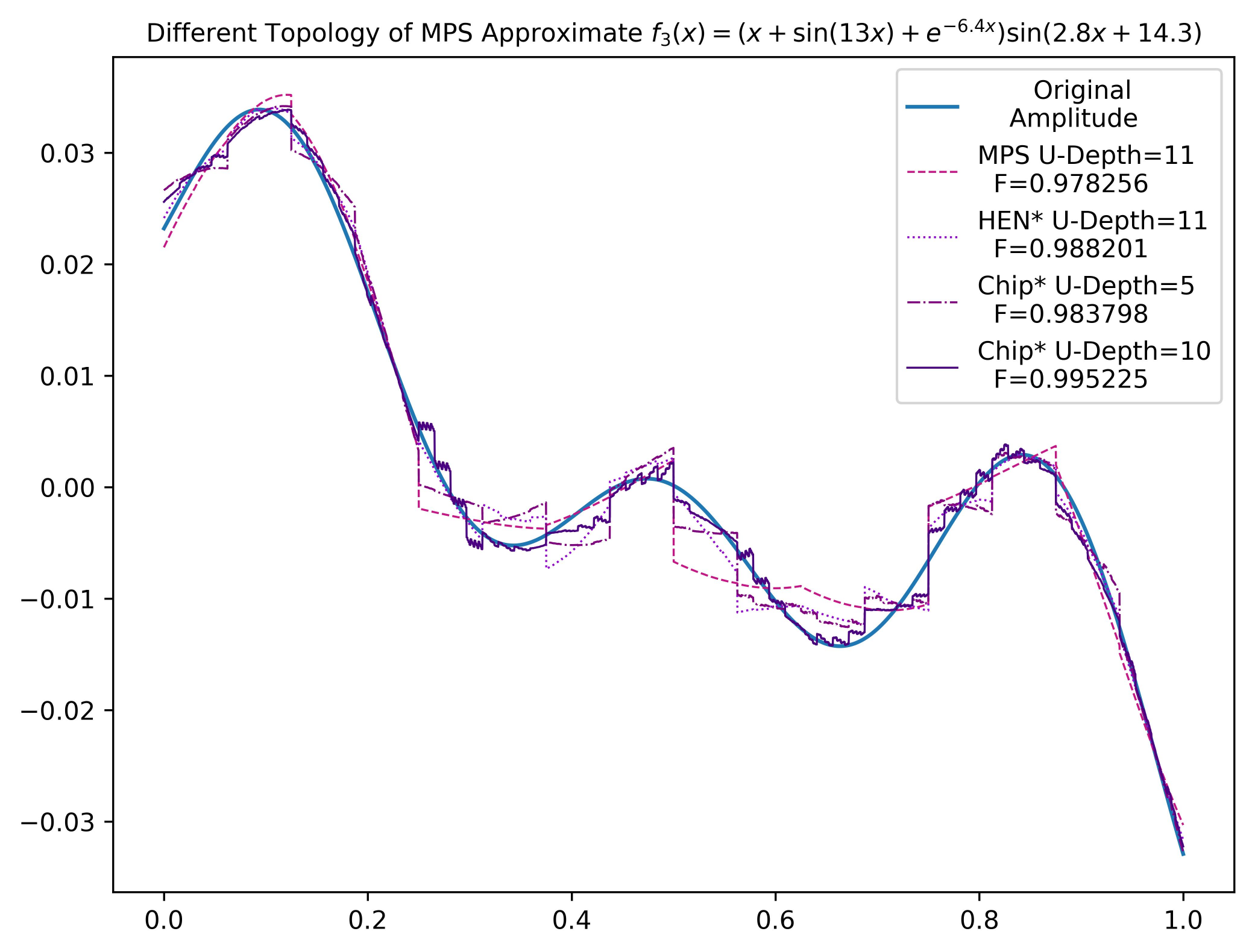}\\
		\includegraphics[width=0.3\textwidth]{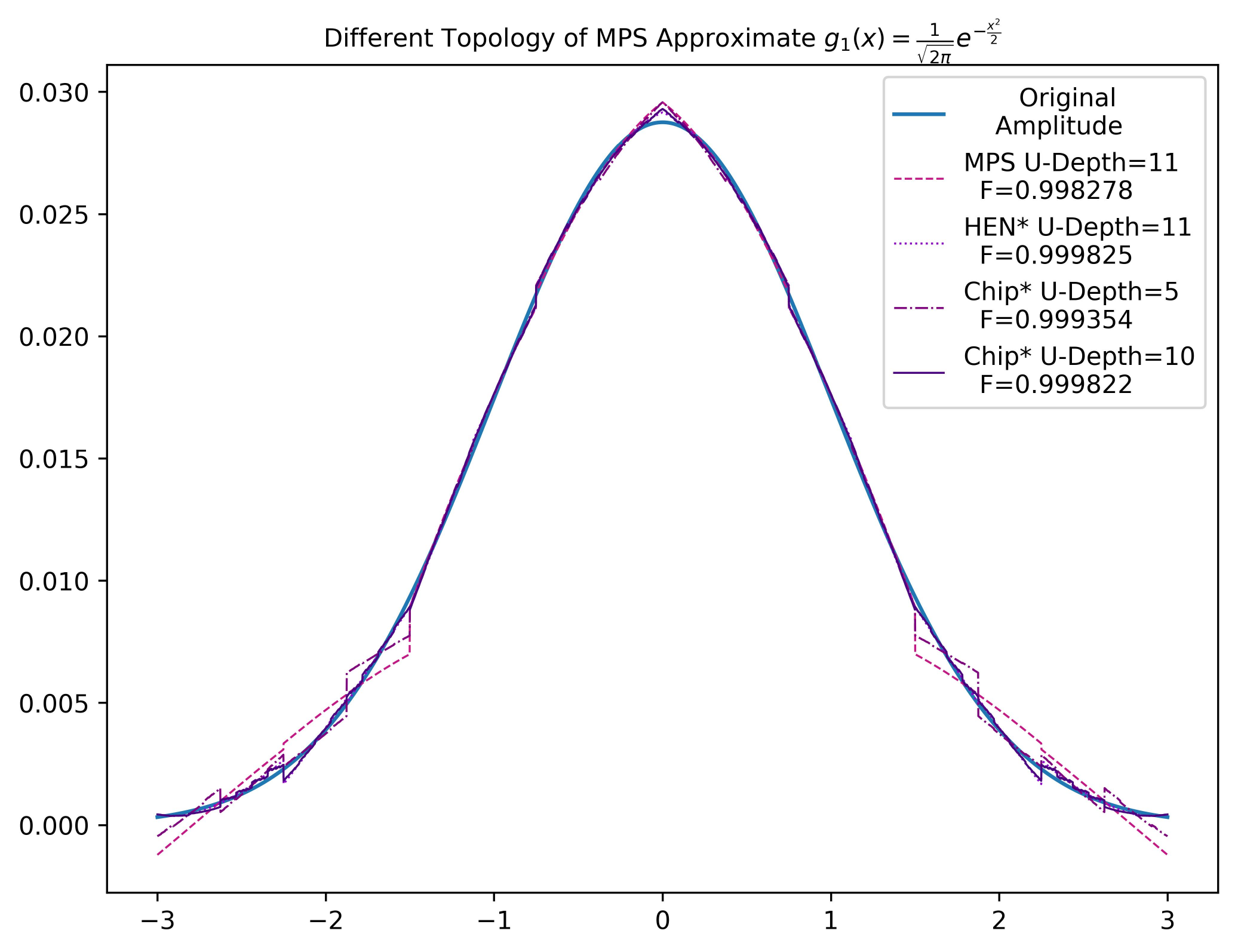}
		\includegraphics[width=0.3\textwidth]{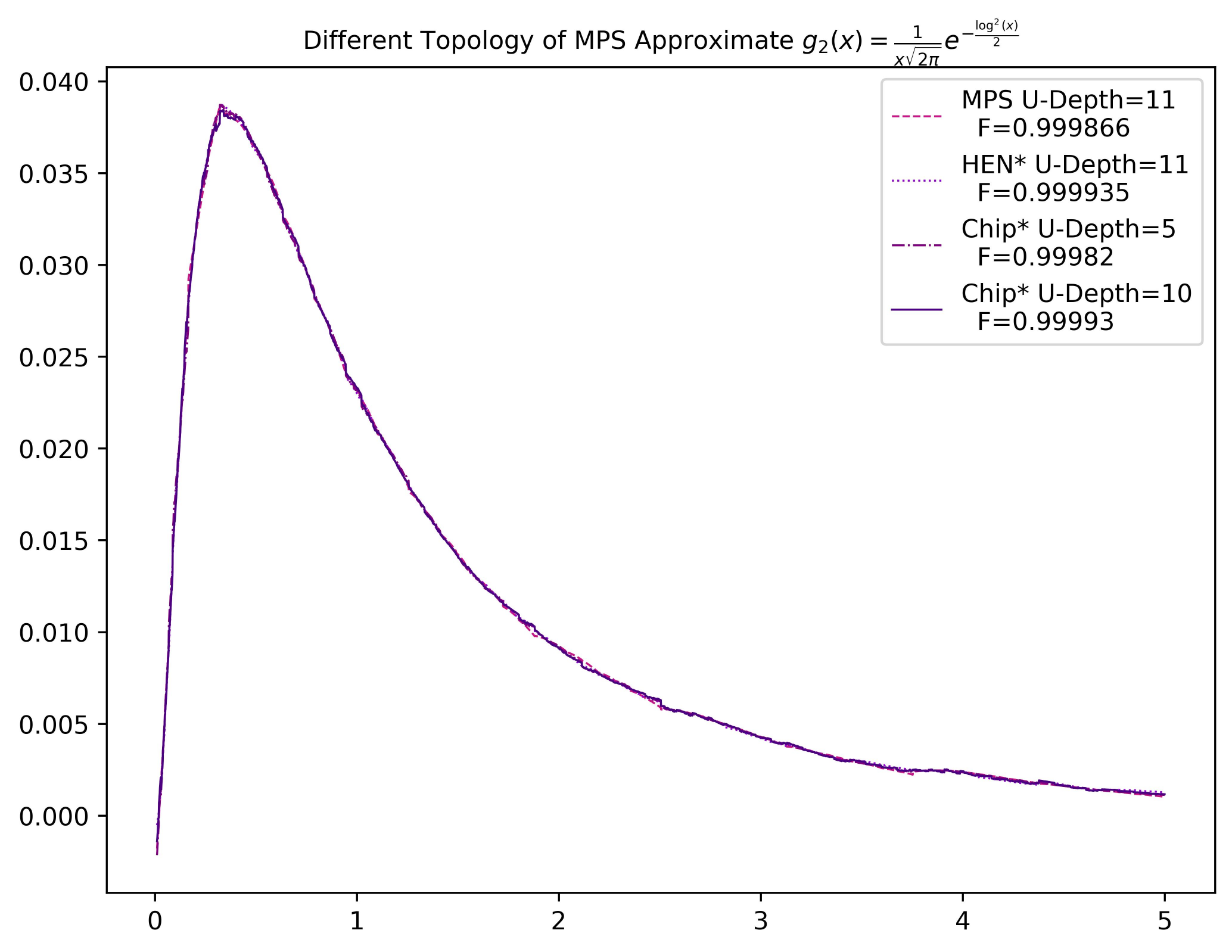}
		\includegraphics[width=0.3\textwidth]{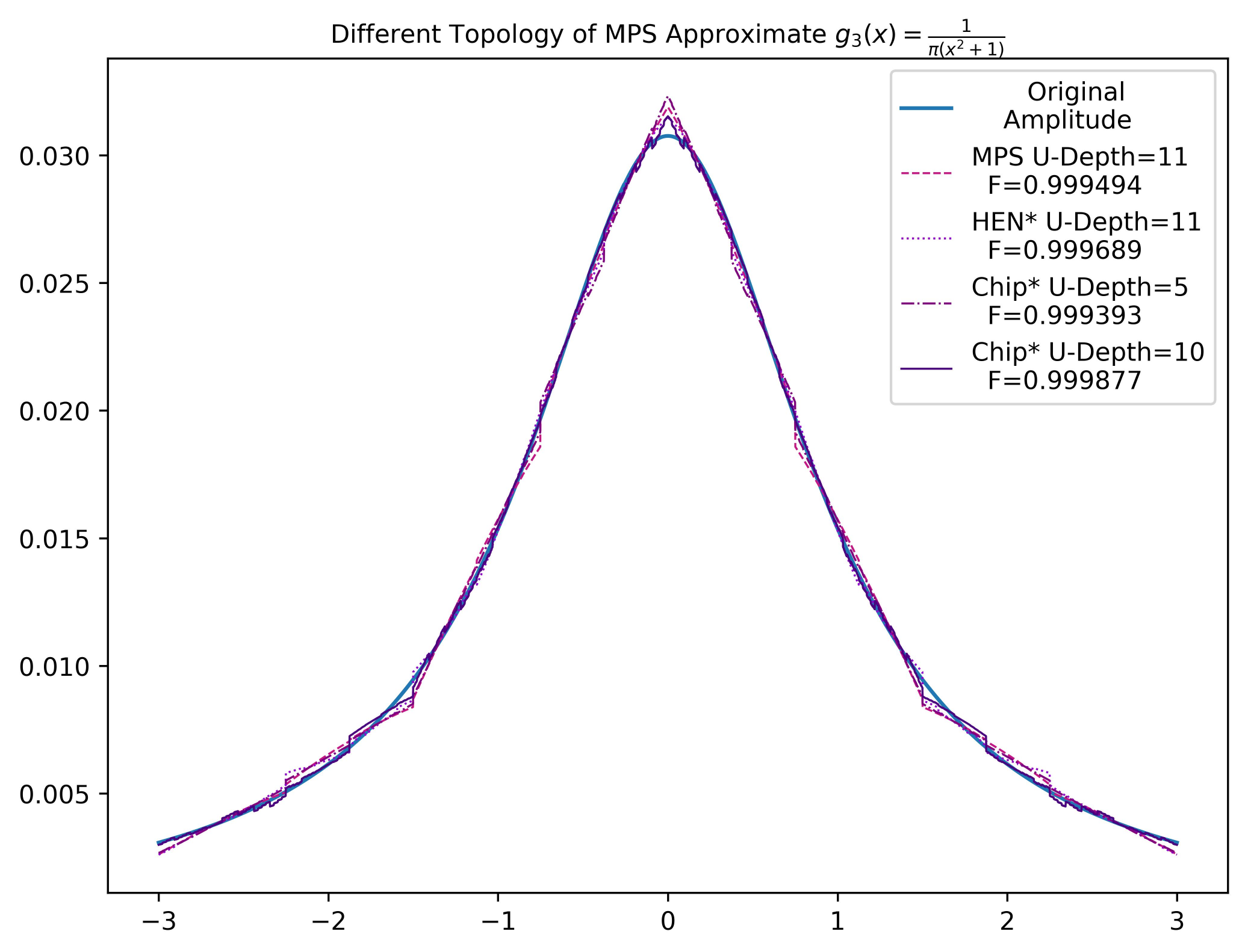}
		\caption{Intuitive comparison of the amplitude approximation results for the traditional MPS and the chip-adapted IMPS at different circuit depths, illustrating the differences between the original distributions and the prepared amplitude distributions.}
		\label{toptop}
	\end{figure}

In terms of characterizing chip adaptability, we conducted a benchmark comparison using the six functions described earlier, with the resulting infidelity comparison shown in Fig.~\ref{Chiptop}. Our scheme outperforms the traditional method in most cases at a U-depth of 5 and comprehensively surpasses its fidelity when using 2 layers (i.e., a U-depth of 10). Additionally, Fig.~\ref{toptop} provides a detailed visual comparison of the distribution fitting results.

	\section{Discussion and Conclusion}
	
	Our IMPS framework enables efficient quantum state preparation with $\mathcal{O}(\log n)$ circuit depth, validated on bounded-rank MPS functions and common distributions. Then we extended this method to a connectivity example in a common superconducting quantum computer topology, achieving reduced circuit depth. However, optimizing circuit depth for arbitrary topologies remains challenging, with reinforcement learning yielding promising preliminary results but lacking formal mathematical solutions. We decompose two-qubit gates in the MPS framework into two CNOT gates and single-qubit gates, lowering hardware resource demands. Extensive hardware experiments are required to substantiate this work, alongside further optimization to mitigate noise and decoherence for enhanced framework reliability and performance.
%	\clearpage
	%\begin{spacing}{1.2}
	\bibliographystyle{unsrt} 
	\bibliography{IMPS}
	%\end{spacing}
	
%	\clearpage
	\appendix
	\section{Proof of 4-Order Unitary Decomposition Theorem~\ref{theorem2}}
	\label{appendix-proof}
	
	In this appendix, we present a detailed proof of Theorem~\ref{theorem2}. Before presenting the detailed proof, we note that in MPS framework, a fourth-order unitary matrix applied to qubits $a$ and $b$ is designed to reduce the amplitude of qubit $a$ to near zero, thus achieving approximate disentanglement of qubit $a$. We will demonstrate that this unitary matrix can be realized using two CNOT gates and several single-qubit gates. To facilitate the formal proof, we introduce four essential lemmas.
	\begin{lemma}[Vidal~\cite{vidal2003universal} Theorem2]
		\label{lemma1}
		Any fourth-order unitary matrix can be decomposed into the form 
		\begin{align}[U_{1}\otimes U_{2}]\exp\{\mathrm{i}(aXX+bYY+cZZ)\}[U_{3}\otimes U_{4}],\end{align}
		$a,b,c \in \mathbb{R}$, $U_{1},U_{2},U_{3},U_{4} \in \mathfrak{u}(2)$ and when $abc=0$, the corresponding circuit can be implemented using at most two CNOT gates and several single-qubit gates.
	\end{lemma}
	\begin{lemma}
		\label{lemma2}
		For $\forall U\in \mathfrak{u}(4)$, if matrix $U[Z\otimes I]U^{\dagger}$ can be implemented using two CNOT gates and several single-qubit gates, then matrix $U_{2cx}$~(\ref{u2cx}) can be implemented in the same manner. Here $Z$ means pauli-Z gate matrix on one qubit.
	\end{lemma}
	\begin{proof}
		Clearly, matrix $\begin{bmatrix}
			C & S \\[0.5ex]
			S & -C
		\end{bmatrix}$ is real and symmetric, and thus can be diagonalized by an orthogonal matrix $P$. The characteristic polynomial of the matrix is computed to be the root of \begin{align}[(\lambda -c_{1})(\lambda +c_{1})-s_{1}^{2}][(\lambda -c_{2})(\lambda +c_{2})-s_{2}^{2}]=0,\end{align} yielding eigenvalues $1, 1, -1, -1$, the elements $c_{1},c_{2},s_{1},s_{2}$ are derived from the diagonal element of matrix $C,S$ at the corresponding position. Consequently, there exists an orthogonal matrix $P$ such that \begin{align}PAP^\mathrm{T} = \left[\begin{matrix}
				1&0&0&0\\[0.5ex]
				0&1&0&0\\[0.5ex]
				0&0&-1&0\\[0.5ex]
				0&0&0&-1
			\end{matrix}\right]=Z\otimes I.\end{align} Setting $P\begin{bmatrix}
			A & O \\[0.5ex]
			O & B
		\end{bmatrix} = U^{\dagger}$, we have \begin{align}U_{2cx}=\begin{bmatrix}
				A^{-1} & O \\[0.5ex]
				O & B^{-1}
			\end{bmatrix}\begin{bmatrix}
				C & S \\[0.5ex]
				S & -C
			\end{bmatrix}\begin{bmatrix}
				A & O \\[0.5ex]
				O & B
			\end{bmatrix}=\begin{bmatrix}
				A^{-1} & O \\[0.5ex]
				O & B^{-1}
			\end{bmatrix}
			P^\mathrm{T}[Z\otimes I]P
			\begin{bmatrix}
				A & O \\[0.5ex]
				O & B
			\end{bmatrix}=U[Z\otimes I]U^{\dagger},\end{align}the problem reduces to the form of matrix $U[Z\otimes I]U^{\dagger}$, which, by the given conditions, can be implemented using two CNOT gates and several single-qubit gates.
	\end{proof}
	Through the proof of Lemma~\ref{lemma2}, we establish that demonstrating the realizability of matrix $U[Z\otimes I]U^{\dagger}$ using two two-qubit gates and several single-qubit gates suffices to complete the proof of Theorem~\ref{theorem2}. Building upon this reformulation, we proceed to construct Lemma~\ref{lemma3} and Lemma~\ref{lemma4} as follows.
	\begin{lemma}
		\label{lemma3}
		For $\forall U\in \mathfrak{u}(4)$, matrix of $U[Z\otimes I]U^{\dagger}$ can be decomposed to $P\cdot \exp(\mathrm{i}\Theta)\cdot Q^{\mathrm{T}}$, where $\Theta=\{\theta_{0},\theta_{1},\theta_{2},\theta_{3}\}$ is a set of angles, if $\alpha \in \Theta$, we have $-\alpha \in \Theta$.
	\end{lemma}
	\begin{proof}
		First we define $K=U[Z\otimes I]U^{\dagger}$, and $K_{R} =\dfrac{K+K^{*}}{2}$,$K_{I} =\dfrac{K-K^{*}}{2\mathrm{i}}$. We have \begin{align}
			K_{R}^{\mathrm{T}} = \frac{K^{\mathrm{T}}+K^{\dagger}}{2}=\frac{K^{*}+K}{2}=K_{R}.
		\end{align}
		So there exist $P_{1} \in \mathfrak{so}(4)$, $P_{1}D_{R}P_{1}^{\mathrm{T}}=K_{R}$, where $D_{R}$ is a real diagonal matrix. 
		
		Let $\mathrm{sgn}(D_{R})$ denote the sign function of a diagonal matrix $D_{R}$, defined as a diagonal matrix whose diag entries are either $+1$ or $-1$. The notation $\mathrm{abs}(D_{R})$ denotes the operation of taking the absolute value of each element in matrix $D_{R}$. Specifically, for each positive diagonal entry of $D$, the corresponding entry in $\mathrm{sgn}(D)$ is $+1$, and for each negative diagonal entry, it is $-1$. From Tucci~\cite{tucci2005introduction} and the proof of its Theorem4, the matrix of $P_{1}^{K} K_{I} P_{1}\mathrm{sgn}(D_{R})$ and $\mathrm{abs}(D_R)$ are commutative, here $P_{1}^{K} K_{I} P_{1}\mathrm{sgn}(D_{R})$ is also real symmetric matrix. 
		
		Therefore, there exists another orthogonal matrix $P_{2}$ that simultaneously diagonalizes both matrices $P_{1}^{\mathrm{T}} K_{I} P_{1}\mathrm{sgn}(D_{R})$ and $\mathrm{abs}(D_R)$, i.e. $P_{2}^{\mathrm{T}}P_{1}^{\mathrm{T}}\cdot K_{I} \cdot P_{1}\mathrm{sgn}(D_{R})P_{2}=\Lambda_{I}$, and $P_{2}^{\mathrm{T}}P_{1}^{\mathrm{T}}\cdot T_{R} \cdot P_{1}\mathrm{sgn}(D_{R})P_{2}=\Lambda_{R}$, here $\Lambda_{R}$ and $\Lambda_{I}$ both diagonal matrix.
		
		Let $P=P_{1}P_{2}$, $Q=P_{1}\mathrm{sgn}(D_{R})P_{2}$, we have \begin{align}
			P^{\mathrm{T}}KQ=P^{\mathrm{T}}(K_R+\mathrm{i}K_I)Q=\Lambda_R+\mathrm{i}\Lambda_I=\exp(\mathrm{i}\cdot \mathrm{diag}\{\theta_{0},\theta_{1},\theta_{2},\theta_{3}\}),
		\end{align}
		equivalently, we have 
		have \begin{align}
			K=P\cdot \exp(\mathrm{i}\cdot \mathrm{diag}\{\theta_{0},\theta_{1},\theta_{2},\theta_{3}\})\cdot Q^{\mathrm{T}}=P\cdot \mathrm{\exp}(\mathrm{i}\Theta)\cdot Q^{\mathrm{T}}.
		\end{align}
		
		From the perspective of determinants, we have $\det(K) = 1$, and $P, Q \in \mathfrak{o}(4)$. if $\det(P)=-1$, let $P\cdot \mathrm{diag}(1, 1, 1, -1)$ replace $P$, and if $\det(Q) = -1$, the same substitution is applied as for $P$. When exactly one of $P$ or $Q$ has a determinant of $-1$, the value of $\theta_{3}$ in $D$ must be adjusted value of $\pi$. In any case, we have $P, Q \in \mathfrak{so}(4)$ such that $K=P\cdot \mathrm{\exp}(\mathrm{i}\Theta)\cdot Q^{\mathrm{T}}$.
		
		Now let's consider one of the angles $\alpha \in \Theta$, which is equivalent to $\mathrm{det}(K-\exp(\mathrm{i}\alpha)PQ^{\mathrm{T}})=0$
		,then we take the conjugate transposition of the matrix, we have \begin{align}
			\mathrm{det}(K-\exp(\mathrm{i}\alpha)PQ^{\mathrm{T}})=\mathrm{det}(K-\exp(-\mathrm{i}\alpha)QP^{\mathrm{T}})=\mathrm{det}(K-\exp(-\mathrm{i}\alpha)PQ^{\mathrm{T}})=0,
			\label{eq16}
		\end{align}
		the second equation holds because 
		\begin{align}
			QP^{\mathrm{T}}=P_{1}\mathrm{sgn}(D_{R})P_{2}\cdot P_{2}^{\mathrm{T}}P_{1}^{\mathrm{T}}=P_{1}\mathrm{sgn}(D_{R})P_{1}^{\mathrm{T}}=P_{1}P_{2} \cdot P_{2}^{\mathrm{T}}\mathrm{sgn}(D_{R}) P_{1}^{\mathrm{T}}=PQ^{\mathrm{T}}.
		\end{align}
		The establishment of Equation~\ref{eq16} mean that
		when $\alpha \in \Theta$ is established, $-\alpha \in \Theta$ is also established. Then we completed the proof.
	\end{proof}
	\begin{lemma}
		\label{lemma4}
		For $\forall U\in \mathfrak{u}(4)$, matrix of $U[Z\otimes I]U^{\dagger}$ can be implemented using two CNOT gates and several single-qubit gates on quantum circuit.
	\end{lemma}
	\begin{proof}
		The magic base matrix $M$ is defined in Tucci~\cite{tucci2005introduction}.
		\begin{align}
			M=\frac{1}{\sqrt{2}}\left[\begin{matrix}
				1&0&0&\mathrm{i}\\[0.5ex]
				0&\mathrm{i}&1&0\\[0.5ex]
				0&\mathrm{i}&-1&0\\[0.5ex]
				1&0&0&-\mathrm{i}
			\end{matrix}\right].
		\end{align}
		From Lemma~\ref{lemma3} we have follow decomposition $M^{\dagger}U[Z\otimes I]U^{\dagger}M=P\cdot \exp(\mathrm{i}\Theta)\cdot Q^{\mathrm{T}}$, and $U[Z\otimes I]U^{\dagger}$ can be written as $M P M^{\dagger} \cdot M\exp(\mathrm{i}\Theta) M^{\dagger}\cdot MQ^{\mathrm{T}}M^{\dagger}$. Then from Tucci~\cite{tucci2004qc} we know that $M P M^{\dagger}$ and $MQ^{\mathrm{T}}M^{\dagger}$ can be realized by $\mathfrak{u}(2)\otimes\mathfrak{u}(2) $. The part of $M\exp(\mathrm{i}\Theta) M^{\dagger}$ can be write as $\exp\{\mathrm{i}(\omega_{0}I\otimes I+\omega_{1}X\otimes X+\omega_{2}Y\otimes Y+\omega_{3}Z\otimes Z)\}$ by a matrix multiplication of $\Gamma$, i.e. $\Gamma^{-1}\Theta=\Omega=[\omega_{0},\omega_{1},\omega_{2},\omega_{3}]^{\mathrm{T}}$, where
		\begin{align}
			\Gamma=\left[\begin{matrix}
				1&1&-1&1\\
				1&1&1&-1\\
				1&-1&-1&-1\\
				1&-1&1&1
			\end{matrix}\right],
		\end{align}
		and $\Gamma^{-1}=\frac{\Gamma^{\mathrm{T}}}{4}$.By Lemma~\ref{lemma3} the elements in $\Theta$ are pairwise opposites. By enumeration, we have $\omega_0=0$, and at least one of $\omega_1,\omega_2,\omega_3$ is 0. Then, by Lemma~\ref{lemma1}, the part of $M\exp(\mathrm{i}\Theta) M^{\dagger}$ can be realized by two CNOT gates and several single-qubit gates, here we completed the proof of Theorem~\ref{theorem2}. 
	\end{proof}

	\end{document}